\documentclass[apj,twocolumn,floatfix,twocolappendix,numberedappendix,appendixfloats,colorlinks]{openjournal}
\usepackage{booktabs}
\usepackage{soul}
\usepackage[dvipsnames]{xcolor}
\usepackage{amsmath}
\usepackage{textcomp,gensymb}
\usepackage{orcidlink}
\usepackage{tabularx}
\hypersetup{linkcolor=red,citecolor=Blue,filecolor=cyan,urlcolor=magenta}
\usepackage[T1]{fontenc}
\usepackage{rotating}

\renewcommand{\baselinestretch}{1.05} 
\newcommand{\acent}{$\mathbf{x_0}$}
\newcommand{\avgsnr}{$\langle$SNR$\rangle$}
\newcommand{\arms}{A_{\textrm{RMS}}}
\newcommand{\armssq}{A^2_{\textrm{RMS}}}
\newcommand{\armsres}{A_{\textrm{RMS,0.1\arcsec}}}
\newcommand{\armspc}{A_{\textrm{RMS,200\,pc}}}
\newcommand{\armsx}{A_{\textrm{RMS,x\arcsec}}}
\newcommand{\armsxpc}{A_{\textrm{RMS,x\,pc}}}
\newcommand{\acas}{A_{\rm{CAS}}}

\newcommand{\new}{\textcolor{black}}

\shorttitle{RMS asymmetry: a robust metric of galaxy shapes}
\shortauthors{Sazonova et al.}
\graphicspath{{./}{}}

\begin{document}

\title{RMS asymmetry: a robust metric of galaxy shapes in images with varied depth and resolution.}

\author{\vspace{-1.3cm}Elizaveta Sazonova\,\orcidlink{0000-0001-6245-5121}$^{1,2}$}
\author{Cameron R. Morgan\,\orcidlink{0009-0009-2522-3685}$^{1,2}$}
\author{Michael Balogh\,\orcidlink{0000-0003-4849-9536}$^{1,2}$}
\author{Katherine Alatalo\,\orcidlink{0000-0002-4261-2326}$^{3,4}$}
\author{Jose A. Benavides\,\orcidlink{0000-0003-1896-0424}$^{5}$}
\author{Asa Bluck\,\orcidlink{0000-0001-6395-4504}$^{6}$}
\author{Sarah Brough\,\orcidlink{0000-0002-9796-1363}$^{7}$}
\author{Innocenza Busa\,\orcidlink{0000-0003-2876-3563}$^{8}$}
\author{Ricardo Demarco\,\orcidlink{0000-0003-3921-2177}$^{9}$}
\author{Darko Donevski\,\orcidlink{0000-0001-5341-2162}$^{10,11}$}
\author{Miguel Figueira\,\orcidlink{0000-0002-9068-6215}$^{10,12}$}
\author{Garreth Martin\,\orcidlink{0000-0003-2939-8668}$^{13}$}
\author{James R Mullaney\,\orcidlink{0000-0002-3126-6712}$^{14}$}
\author{Vicente Rodriguez-Gomez\,\orcidlink{0000-0002-9495-0079}$^{15}$}
\author{Javier Román\,\orcidlink{0000-0002-3849-3467}$^{16,17}$}
\author{Kate Rowlands\,\orcidlink{0000-0001-7883-8434}$^{18,4}$\vspace{0.5cm}}

\affiliation{$^{1}$Waterloo Centre for Astrophysics, University of Waterloo, Waterloo, ON, N2L 3G1 Canada}
\affiliation{$^{2}$Department of Physics and Astronomy, University of Waterloo, Waterloo, ON N2L 3G1, Canada}
\affiliation{$^{3}$Space Telescope Science Institute, 3700 San Martin Dr, Baltimore, MD 21218, USA}
\affiliation{$^{4}$William H. Miller III Department of Physics and Astronomy, Johns Hopkins University, Baltimore, MD 21218, USA}
\affiliation{$^{5}$Department of Physics and Astronomy, University of California, Riverside, 900 University Avenue, Riverside, CA 92521, USA}
\affiliation{$^{6}$Stocker AstroScience Center, Department of Physics, Florida International University, Miami, FL, 33199, USA}
\affiliation{$^{7}$School of Physics, University of New South Wales, NSW 2052, Australia}
\affiliation{$^{8}$INAF, Osservatorio Astrofisico di Catania, Via S. Sofia 78, I-95123 Catania, Italy}
\affiliation{$^{9}$Institute of Astrophysics, Facultad de Ciencias Exactas, Universidad Andr\'es Bello, Sede Concepci\'on, Talcahuano, Chile}
\affiliation{$^{10}$National Centre for Nuclear Research, Pasteura 7, 02-093 Warsaw, Poland}
\affiliation{$^{11}$SISSA, Via Bonomea 265, 34136 Trieste, Italy}
\affiliation{$^{12}$Max-Planck-Institut für Radioastronomie, Auf dem Hügel 69, 53121, Bonn, Germany}
\affiliation{$^{13}$School of Physics and Astronomy, University of Nottingham, University Park, Nottingham NG7 2RD, UK}
\affiliation{$^{14}$Department of Physics and Astronomy, University of Sheffield, Sheffield S3 7RH, UK}
\affiliation{$^{15}$Instituto de Radioastronom\'ia y Astrof\'isica, Universidad Nacional Aut\'onoma de M\'exico, A.P. 72-3, 58089 Morelia, Mexico}
\affiliation{$^{16}$Departamento de Astrofísica, Universidad de La Laguna, E-38206, La Laguna, Tenerife, Spain}
\affiliation{$^{17}$Instituto de Astrofísica de Canarias, c/ Vía Láctea s/n, E-38205, La Laguna, Tenerife, Spain}
\affiliation{$^{18}$AURA for ESA, Space Telescope Science Institute, 3700 San Martin Drive, Baltimore, MD, USA}

\begin{abstract}

Structural disturbances, such as galaxy mergers or instabilities, are key candidates for driving galaxy evolution, so it is important to detect and quantify galaxies hosting these disturbances spanning a range of masses, environments, and cosmic times. Traditionally, this is done by quantifying the asymmetry of a galaxy as part of the concentration-asymmetry-smoothness system, $\acas{}$, and selecting galaxies above a certain threshold as merger candidates. However, in this work, we show that $
\acas{}$, is extremely dependent on imaging properties -- both resolution and depth -- and thus defining a single $\acas{}$ threshold is impossible. We analyze an alternative root-mean-squared asymmetry, $\arms$,  and show that it is independent of noise down to the average SNR per pixel of 1. However, both metrics depend on the resolution. We argue that asymmetry is, by design, always a scale-dependent measurement, and it is essential to define an asymmetry at a given physical resolution, where the limit should be defined by the size of the smallest features one wishes to detect. We measure asymmetry of a set of $z\approx0.1$ galaxies observed with \textit{HST}, HSC, and SDSS, and show that after matching the resolution of all images to 200~pc, we are able to obtain consistent $\armspc$ measurements with all three instruments despite the vast differences in the original resolution or depth. We recommend that future studies use $\armsxpc$ measurement when evaluating asymmetry, where $x$ is defined by the physical size of the features of interest, and is kept consistent across the dataset, especially when the redshift or image properties of galaxies in the dataset vary.

\end{abstract}

\keywords{Galaxy structure (622) --- Galaxy morphology (582) --- astronomy image processing (2306)}

\maketitle

\section{Introduction} \label{sec:intro}

There is a strong bimodality among galaxies in the local Universe, with two dominant populations: blue star-forming and red quiescent galaxies \citep[e.g.,][]{Baldry2004,Balogh2004,Taylor2015}. Studies into the origin of this bimodality show that the fraction of star-forming galaxies of a given mass decreases with cosmic time while the fraction of quiescent ones increases, indicating evolution \citep{Bell2004, Arnouts2007, Faber2007}. Their stellar structure is also bimodal: while blue galaxies are typically ``late-type'' disks with prominent spiral arms, most quiescent galaxies are ``early-type'' with a smooth spheroidal structure \citep[e.g.,][]{Strateva2001,Schawinski2014}, with rare exceptions \citep[e.g.,][]{Masters2010}. The existence of a \textit{morphology} bimodality implies that the structure of galaxies must also evolve during this transition. However, the relative timescales of the two changes, or the causal relationship between star formation and structure are still disputed.

As a result, the processes driving this evolution are also debated, and likely vary from one galaxy to another, depending on many factors such as mass or environment \citep[e.g.,][]{Baldry2006,Peng2010}. A combination of different quenching mechanisms can be responsible for shutting down star formation in a galaxy, temporarily or permanently. Typically, they act in one of three ways: by 1) removing the gas supply, as do supernovae-driven winds \citep[e.g.,][]{Heckman1990,Leroy2015}, winds originating from the active galactic nuclei \citep[e.g.,][]{Cicone2014}, or ram pressure in galaxy clusters \citep{Gunn1972,Abadi1999,Poggianti2017}; 2) triggering a period of rapid star formation that quickly consumes the gas, for example from bar instabilities \citep[e.g.,][]{Masters2012} or radial gas inflows following a merger \citep[e.g.,][]{Hopkins2009}; or 3) by preventing the gas from cooling enough to form stars \citep[e.g., through heating by the central black hole or stabilization in the bulge;][]{Benson2003,Martig2009,Werner2019}. This list is by far not exhaustive, and many of these mechanisms can (and likely do) act in tandem, or trigger other mechanisms. For example, a galaxy merger may both trigger a starburst and feed the central black hole, leading to rapid gas consumption and its eventual expulsion in massive outflows, while black hole feedback may be needed to ensure star formation does not resume \citep[e.g.,][]{Ellison2013,Piotrowska2022,Bluck2023}.

Both the gas properties of a galaxy and its structure change as it evolves, however it is still unclear which change occurs first. Therefore, studying the structural impacts of different quenching pathways is crucial. Most physical processes affect the observed morphology in some way: mergers can induce tidal features such as shells or tails \citep[e.g.,][]{Toomre1972,Schweizer1980}, outflows can be detected as asymmetric features in gas and dust maps \citep[e.g.,][]{Quinn1984}, while bars are identifiable as axisymmetric perturbations \citep[e.g.,][]{Ciambur2016,Kruk2018}. Detecting disturbances in galaxy imaging can therefore help us identify galaxies which are currently quenching, and provide insight into the ongoing physical mechanisms.

Doing this in a fast, accurate, and an automated way allows studying large sets of galaxies, and quickly leveraging large surveys, such as Sloan Digital Sky Survey \citep[SDSS;][]{sdss,sdss4}. Upcoming surveys, such as Rubin Observatory's Legacy Survey of Space and Time \citep[LSST;][]{lsst} and the Euclid survey \citep{euclid}, will provide an unprecedented dataset of billions of galaxies, including low surface brightness ones that were challenging to study before. Thus, a set of structural measurements are needed to detect and quantify structural disturbances for these large datasets. 

A classic one is the measurement of a galaxy's asymmetry \citep[e.g.,][]{Schade1995,Abraham1996}. Currently, the most widely used metric is the asymmetry defined as part of the concentration-asymmetry-smoothness parameter system \citep[CAS;][]{Conselice2000,Conselice2003}, hereafter $\acas{}$, calculated by rotating the image by 180$\degree$, finding the absolute residual between the original and the rotated images, and computing the ratio of this residual to the total flux. This metric is sensitive to bright regions that deviate from an otherwise symmetric profile, and has traditionally been used to detect major mergers \citep[e.g.,][]{Conselice2008,Lotz2011}. Recently, $\acas{}$ has been shown to detect a broader set of features: disrupted dust structure \citep[e.g.,][]{Sazonova2021}, clumpy star formation \citep[e.g.,][]{Ferreira2022a}, outflows in H\textsc{I} maps \citep{Deg2023, Holwerda2011}, and more.

However, the $\acas{}$ measurement is highly dependent on both image resolution and signal-to-noise ratio \citep[e.g.,][]{Lotz2004,Giese2016,Bottrell2019,Bignone2020,Thorp2021}. Therefore, comparing measurements of asymmetry across datasets is challenging. $\acas{}$ measured on noisy or poorly resolved images can miss important features, so using this metric alone can lead to underestimating the fractions of merging or disturbed galaxies \citep[e.g.,][]{Snyder2019,Sazonova2021,Wilkinson2024}. Newer approaches have been developed that address some of these issues: for example, shape asymmetry, which is insensitive to image resolution \citep{Pawlik2016}; neural networks \citep[e.g.,][]{DominguezSanchez2018,Walmsley2022,Desmons2023}, or visual classification schemes \citep[e.g., Galaxy Zoo;][]{Lintott2008, Willett2013}.

Despite its strong dependence on image quality, the traditional $\acas{}$ measurement is advantageous in that it is easy to interpret (unlike many machine learning methods) and is relatively fast to compute on large datasets. Moreover, $\acas{}$ is  widely used to this day and thus enables easy comparisons between studies, provided similar imaging quality or source redshifts. To make $\acas{}$ more robust, various methods to correct for the impact of noise or resolution on $\acas{}$ have been proposed \citep{Shi2009,Wen2016,Thorp2021,Yu2023}. However, while the numerical performance of $\acas{}$ has been tested, few studies have tested the mathematical definition of $\acas{}$ and its \textit{expected} behavior to varying noise or resolution. \cite{Thorp2021} show that the background contribution is always overestimated in the $\acas{}$ measurement and provide a correction, while \cite{Conselice2000} qualitatively note that asymmetry depends on resolution. 

In this paper, we mathematically show that $\acas{}$ depends strongly on background due to the absolute-valued definition, while the resolution dependence arises from both the pixelation and the smearing by the point spread function (PSF). Similarly to \cite{Deg2023}, we propose to instead use the root-mean-square asymmetry, $\arms$, as initially considered in \cite{Conselice1997} and \cite{Conselice2000}. 

We then perform a series of simulations using \textsc{GalSim} \citep{galsim} and test the dependence of both $\acas{}$ and $\arms$ on image noise and resolution. We further test these measurements on a pilot dataset of 27 galaxies studied in \cite{Sazonova2021}, which span a range of morphologies and degrees of disturbance, and are observed with the the Hubble Space Telescope (\textit{HST}), SDSS, and Hyper-Suprime Cam \citep[HSC;][]{hsc}, thereby sampling different image depths and resolution. Finally, we use our results to provide guidelines on which parameter performs best in different experimental set-ups. 

The paper is organized as follows. We describe the \textsc{GalSim} simulations used throughout the paper in Section \ref{sec:sim}. In Section \ref{sec:cas}, we show the original $\acas{}$ definition and demonstrate its mathematical dependence on noise and the PSF size. We also propose a slight modification that alleviates some of the noise dependence. In Section \ref{sec:arms}, we introduce the $\arms$ measurement and show that it has no dependence on noise, but a higher dependence on resolution. We test the performance of these metrics on real data from the SDSS, \textit{HST}, and HSC data for the same set of galaxies in Section \ref{sec:obs}. Finally, we discuss our findings and provide guidelines for the use of the different asymmetry parameters in Section \ref{sec:discussion}.

Throughout this work, we use the Planck 2018 cosmology \citep[H$_0$=67.6, $\Omega_m$=0.310;][]{Planck2018} to convert between angular and physical sizes.

{\renewcommand{\arraystretch}{1.4}
\begin{table}
\label{tab:galsim}
\caption{\textsc{GalSim} parameter distributions}
\begin{tabularx}{\linewidth}{lX}
\hline\hline
Parameter & Distribution \\
\hline
\multicolumn{2}{l}{Galaxy properties} \\
\hline
\textit{r}-mag & Uniform~(11, 18) \\
S\'ersic $n$ & Uniform~(1, 5) \\
$R_{\textrm{eff}}$ [arcsec] & $\new{1 \leq} -1.9\times r\textrm{-mag} + 35 + \mathcal{N}(0, 1.5) \new{ \leq 20}$ \\
Orientation &Uniform~(0, 2$\pi$) \\
Ellipticity & 0 if $n>$~3.5 else Uniform~(0,1) \\
\hline
\multicolumn{2}{l}{Clump properties}\\
\hline
$N$ & Discrete Uniform~(5, 30)\\
Flux fraction  & Log-Uniform~(10$^{-3}$, 0.5) \\
Size [arcsec] & Uniform~(0.1, 3) \\
$\theta$ & Uniform~(0, 2$\pi$) \\
$r$/R$_p$ & Uniform~(0.05, 1)\\
\hline
\multicolumn{2}{l}{Mock image properties}\\
\hline
Pixel scale [arcsec] & Uniform~(0.1, 0.5) \\
PSF FWHM  [arcsec] &  Uniform~(2 $\times$ pixel scale, 3)\\
Sky $\sigma$ [mag/arcsec$^2$] & Uniform~(20, 26)
\end{tabularx}
\end{table}
}

\begin{figure*}
    \centering
    \includegraphics[width=\textwidth]{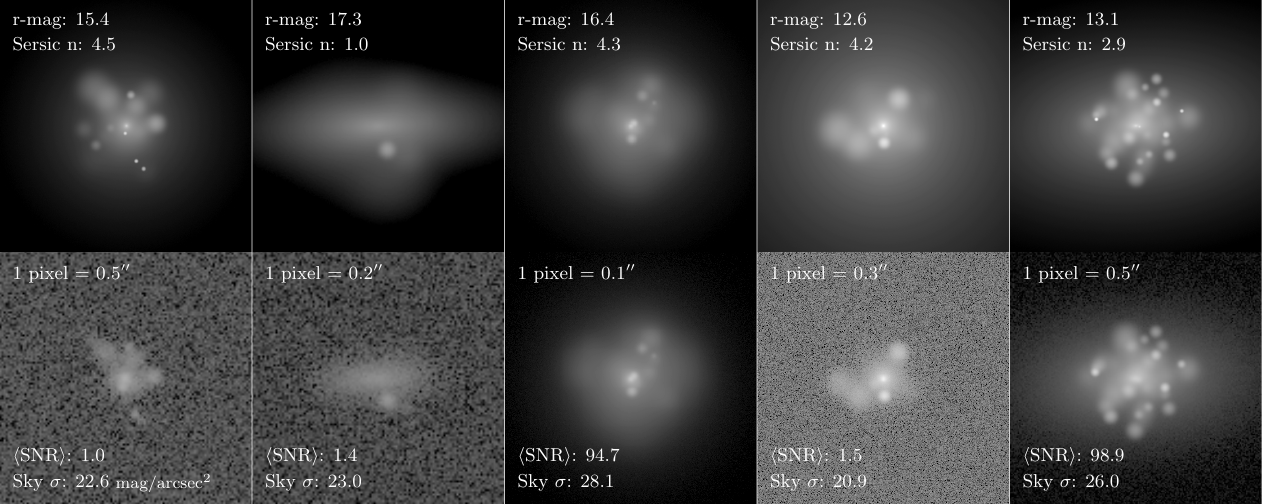}
    \caption{A random sample of baseline simulated galaxies (\textbf{top}) and their mock observations (\textbf{bottom}). Each galaxy is generated as a combination of a S\'ersic profile and a set of Gaussian clumps using a  set of parameters randomly drawn from a distribution summarized in Table \ref{tab:galsim}. The mock image is then generated with a different pixel scale, convolved with a Gaussian PSF, then has a sky background and Poisson noise applied. The pixel scale and 1$\sigma$ image depth are also randomly drawn and shown here for each mock image. The PSF FWHM is chosen to be 3 times the pixel scale to ensure Nyquist sampling. The \avgsnr{} is calculated within a 1.5$R_{p}$ aperture around the source.
    }
    \label{fig:galsim}
\end{figure*}

\section{\textsc{GalSim} mock images}\label{sec:sim}

To test the performance of the asymmetry metrics, we need to compare the asymmetry measured using an imperfect ``observed'' image to the ``true'' asymmetry of a galaxy. Since the ``true'' asymmetry of real galaxies is unknown, we use a simulated dataset, and then test our results on a sample of observed galaxies (Section \ref{sec:obs}). However, even for simulated objects, estimating a ``true'' asymmetry is impossible: since any numerical calculation requires evaluating asymmetry on some discretized pixel grid, any calculation will depend on the chosen resolution. Instead, we simulate pairs of galaxies: a ``baseline'' image, which represents the best-case scenario least affected by observational biases, and an ``observed'' or ``mock'' image with varying resolution and depth.

While we could have used realistic galaxies from cosmological simulations such as Illustris TNG \citep{Pillepich2018,Nelson2019,statmorph,Bottrell2024} or higher-resolution zoom-in simulations \citep[e.g.,][]{Ceverino2014,Hopkins2018}, all numerical simulations still have a limited spatial resolution and do not reproduce clumps smaller than the softening scale. Moreover, while they provide large datasets in general, there is a limit to how many highly asymmetric galaxies we can obtain from simulations since merging galaxies are rare. Instead, we opted for a simpler approach: we produced mock observations using \textsc{GalSim} \citep{galsim}, which produces an image given a set of user-input light distributions and perturbation parameters. This also gave us freedom in varying both the depth and resolution of simulated images. Our approach to use a single S\'ersic profile with Gaussian clumps is essentially the same as that taken to create the Rubin Data Preview 0 dataset \citep{dp0}. 

We defined each galaxy as a single two-dimensional S\'ersic profile \citep{Sersic1963} with a variable total flux, S\'ersic index $n$, orientation, half-light radius $R_{\textrm{eff}}$, and ellipticity. We used the \textsc{petrofit} package \citep{petrofit} to calculate the Petrosian radius \citep[$R_p$,][]{Petrosian1976}, based on the S\'ersic index and half-light radius for future steps. \textsc{petrofit} calculates the ratio of the mean global brightness to the isophotal brightness as a function of radius for a given $n$ and $R_{\textrm{eff}}$, and we picked the radius where this value is equal to 0.2 as $R_p$. For each object, the magnitude, $n$, $R_{\textrm{eff}}$, orientation, and ellipticity were randomly drawn from the distributions in Table \ref{tab:galsim}. The range of values was chosen to produce galaxies consistent with observed z$\approx$0.1 objects used in Section \ref{sec:obs}. The $R_\textrm{eff}$ distribution is derived from estimating the relationship between $R_\textrm{eff}$ and \textit{r}-magnitude and its scatter in SDSS galaxies. 

To create asymmetric features in our simulated galaxies, we added $N$ Gaussian clumps on top of the S\'ersic profile. Each clump is defined using its own size (Gaussian $\sigma$), a flux fraction relative to the total flux of the S\'ersic component, and a location given by an angle $\theta$ and a radius from the galaxy's center, r/R$_p$. All model parameters are again drawn from distributions summarized in Table \ref{tab:galsim}.

Our Gaussian clumps mimic various disturbances seen in real galaxies: small clumps embedded in the disk are most similar to star-forming clumps; off-axis small clumps can be viewed as satellite galaxies, while large and far-removed faint clumps are similar to tidal features. This is clearly an over-simplification of the structure of real galaxies, and does not produce true merger-like structures such as tidal tails. However, the aim of this work is not to devise a metric that is sensitive to a particular disturbance, but to test the performance of existing metrics with regards to the image quality, and these simplistic distributions allow us to test this efficiently. We note that both $\acas{}$ and $\arms$ calculate the total flux in the residual after a rotation (see Sections \ref{sec:cas} and \ref{sec:arms}). Therefore, the values depend on the size and the brightness of disturbed features, but not on their shape or distance from the galactic disk, so the numerical performance on more realistic distributions will remain the same. Since our tests span a wide range of sizes and relative fluxes, we can investigate the performance of the metrics despite our simple galaxy models.

At this stage, the idealized galaxy is described purely by a set of analytical functions and does not suffer from any observational effects. We then construct the ``baseline'' image. We create a blank patch of sky with a field-of-view of 10$R_p$ and a pixel scale of 0.1$\arcsec$. Note that at this stage, we must assume a resolution limit -- we opted for a pixel scale of 0.1$\arcsec$ per pixel ($\sim$200 pc at z=0.1), comparable to that of space-based imaging (e.g., Euclid), as higher-resolution simulations become computationally intensive even with the \textsc{GalSim} approach. Since the image should be Nyquist-sampled, we convolve the model with a Gaussian PSF whose full width at half maximum (FWHM) is three times the size of the pixel scale, similar to typical telescope designs (e.g. SDSS, HSC)\footnote{Note that while Nyquist sampling requires at least 2 resolution elements per PSF FWHM, since pixels are two-dimensional, the requirement along the diagonal becomes $2\sqrt{2}\approx 2.8$ pixels, hence we choose a PSF FWHM that spans 3 pixels}. Thus, the effective resolution of our baseline images is 0.3$\arcsec$. \new{Our simulated galaxies vary from 1$\arcsec$ to 20$\arcsec$ in size, so the baseline images are resolved with 3 to 60 resolution elements.} We then add the galaxy model to the sky patch assuming an SDSS \textit{r}-band bandpass and transmission curve. 

After this, we add a low-level uniform sky background that, based on Poisson statistics, would produce sky noise surface brightness (SB) of 30 mag/arcsec$^2$ in a blank region of the image. The SB is calculated using the pixel-by-pixel 1$\sigma$ standard deviation of the sky. Note that this is a different depth definition than that used in survey design, where either an $N\sigma$ point-source depth or depth within a detection area are often quoted \citep[e.g.,][]{lsst,Roman2020}. When the goal is source detection, these depth metrics are preferable. However, since we are interested in pixel-by-pixel flux variations, we opt for using a pixel-by-pixel surface brightness metric instead since it more directly relates to the features we aim to measure.

Finally, we convert the image to electron counts, apply Poisson noise to the whole image, and subtract the mean sky level to get our baseline image. To create a mock observation, we repeat the same steps with a different resolution and noise level, both drawn from the respective distributions in Table \ref{tab:galsim}. \new{Note that in some cases, the mock galaxies are effectively unresolved: e.g., if $R_\textrm{eff}=1\arcsec$ and the PSF size is $1\arcsec$ or larger.} While we do not set the \avgsnr{} manually, we calculate it by taking the mean SNR within the aperture in which asymmetry is calculated. We run three series of tests, as decribed in detail in Section \ref{sec:cas}: with varying sky surface brightness, varying pixel scale, and varying aperture size. For each series of tests, we generated 1,000 \textsc{Galsim} galaxies. Examples of some baseline and mock observations are shown in Fig. \ref{fig:galsim} (top and bottom rows respectively).

Our simulations and mock observations are tuned to roughly match the imaging of relatively local (z$\sim$0.1) galaxies of different masses; however, the important factors we study are the impact of the spatial resolution and depth. The implications of our results can be easily applied to observations of higher-redshift objects, since higher-redshift objects observed within the same survey have a lower physical resolution and SNR. 

\section{CAS asymmetry}\label{sec:cas}

\begin{figure*}
    \centering
    \includegraphics[width=\linewidth]{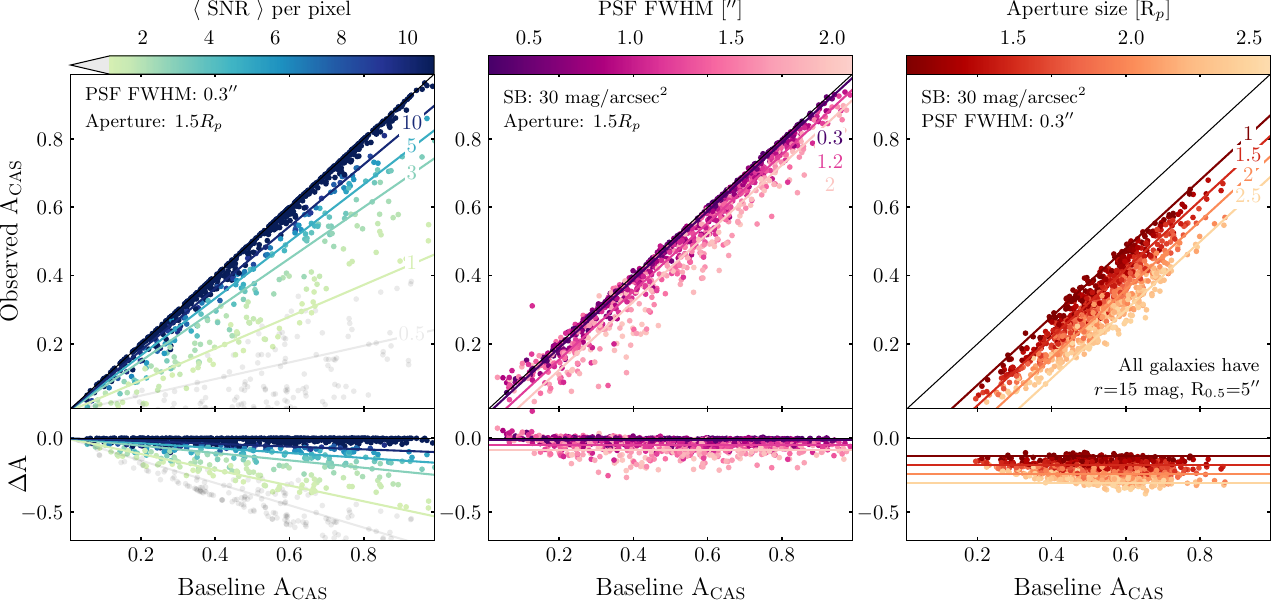}
    \caption{The comparison (\textbf{top}) and the difference (\textbf{bottom}) of $\acas{}$ measured using mock observations (y-axis) and a baseline image (x-axis) for three series of tests on simulated galaxies: varied \avgsnr{} while resolution is fixed to the baseline resolution (\textbf{left}), varied resolution with a fixed baseline depth (\textbf{center}), and varied aperture size with a fixed resolution, depth, and galaxy size and flux (\textbf{right}). The diagonal line shows a 1-to-1 relationship between the measured and baseline asymmetries. Solid lines represent best fits to the dependence of observed $\acas{}$ on the baseline $\acas{}$ and \avgsnr{}, PSF size, and aperture size as described in Sec. \ref{sec:cas}. $\acas{}$ is only weakly dependent on resolution in these tests, but it is significantly underestimated in shallow images compared to the deep baseline ones. This effect depends on the intrinsic asymmetry of the object, where $\acas{}$ of disturbed galaxies is \textit{more} underestimated. Finally, even for a fixed image quality, $\acas{}$ depends on the aperture size, i.e. the number of blank background pixels included in the measurement.}
    \label{fig:a_cas}
\end{figure*}

The most common way to measure asymmetry is to find the absolute difference in flux distributions under a 180$\degree$ rotation, first defined in \cite{Schade1995} and \cite{Abraham1996}. The current implementation of this metric is $\acas{}$, described in \cite{Conselice2000} and \cite{Conselice2003}. It is calculated in the following way. 

An image consists of some light distribution, $I(x,y)$, which is discretized at each pixel as $I_{ij}$. The image is rotated by 180 degrees about the center \acent~to get the flux of the rotated image, $I^{180}_{ij}$. Then asymmetry is calculated in as:

\begin{equation}\label{eq:a_cas}
    \acas{} = \min_{\mathbf{x_0}} \frac{\sum_{ij} |I_{ij} - I^{180}_{ij}|}{\sum_{ij} |I_{ij}|} - \langle \acas{}_{,bg} \rangle
\end{equation}

\noindent where $\langle \acas{}_{,bg} \rangle$ is the average background asymmetry, which we describe in detail below. The sum is performed in an aperture of size $fR_p$ centered on \acent\new{, which minimizes the asymmetry.} The value of $f$ typically ranges between 1 and 2 with most codes adopting $f=1.5$ by default \citep[e.g., the \texttt{statmorph} code;][]{statmorph}, although larger $f$ may be needed for very bulge-dominated galaxies as galaxy flux extends significantly past $2R_p$ \citep{Graham2005}. In some cases, where the underlying distribution is too clumpy to compute $R_p$, instead an aperture derived from another band may be used \citep[e.g.,][]{Holwerda2012,Holwerda2023}. Finally, while most often the aperture is circular, elliptical apertures can be adopted for highly inclined galaxies to minimize the impact of the sky background \citep[e.g., \textsc{Morfometryka};][]{Ferrari2015}.

The crucial steps of this calculation are 1) calculating the background contribution $\acas{}_{,bg}$ and 2) choosing the coordinate \acent~ that minimizes the asymmetry. Both of these steps include several caveats and can be implemented in different ways. \new{In some applications, the minimization step is skipped entirely. This can be beneficial if one can find a physically meaningful asymmetry center, such as in radio astronomy \citep[e.g.,][]{Holwerda2011,Deg2023} or in studies of galaxy clusters \citep[e.g.,][]{DeLuca2021}. In most optical morphology applications, however, this step is necessary, and} the choices for the minimization algorithm can affect the measurement. \new{We discuss different choices and show a potential analytical approach to finding the asymmetry minimum in} Appendix \ref{app:min}. 

Correctly treating the background is even more challenging. The sky may have large-scale asymmetries (e.g., gradients), which need to be subtracted from the image to not affect the measurement. However, even if the large-scale background is perfectly subtracted, the Poisson noise from the sky counts will still affect the calculation as it contributes to the absolute value of the residuals. Therefore, the contribution from the background \textit{noise} needs to be corrected for. Calculating the background asymmetry in the same aperture as the source is impossible, and most codes approximate it from a blank patch of sky in some way, which we discuss in detail in Appendix \ref{app:bg}. However, it is worth noting here that a common approach is to calculate the asymmetry in a blank sky aperture (we use a circular annulus around the source), and then re-normalize the background noise contribution by the ratio of the sizes of the sky and the source apertures ($N_{\rm{sky}}$ and $N_{\rm{ap}}$ respectively). Assuming the image is background-subtracted, the correction is

\begin{equation}\label{eq:a_cas_bg}
    \langle \acas{}_{,bg} \rangle = \frac{N_{\rm{ap}}}{N_{\rm{sky}}}\frac{\sum_{ij}^{sky} |\epsilon_{ij} - \epsilon^{180}_{ij}|}{\sum_{ij} |I_{ij}|}
\end{equation}

\noindent where $\epsilon(x,y)$ is the sky flux distribution and $\epsilon_{ij}$ is its discretized version. In the case of a flat and a subtracted sky background the noise is normally-distributed, i.e. $\langle \epsilon \rangle = 0$.  We note that since the background correction is absolute-valued it is always positive, so it will always decrease the measured asymmetry.

Without any background contribution, $\acas{}$ would theoretically vary between 0 and 2. However, since \acent{} minimizes asymmetry, $\acas{}=2$ is unlikely, and in most cases $0<\acas{}\lessapprox 1$. Typically, a threshold $\acas{}>0.3$ is used to select mergers, although as we show in this work this threshold should depend on the imaging properties. Since $\langle \acas{}_{,bg} \rangle$ is always positive, then if the image is noisy and the galaxy is undisturbed, the resultant $\acas{}$ can be slightly negative \citep[e.g., as shown in][]{statmorph}.

\subsection{Numerical performance}

We investigated the numerical performance of $\acas{}$ using a series of tests, the results of which are shown in Fig. \ref{fig:a_cas}. For each series, we generated 1,000 galaxies with random physical properties as described in Sec. \ref{sec:sim} (except the third series, discussed below), and compared the asymmetry calculated using the baseline (SB=30 mag/arcsec$^2$, 0.1$\arcsec$ pixel scale) and the observed images, shown on the x- and y-axis respectively. We first looked at how observed $\acas{}$ changes with image depth (left panel), then with the resolution (middle panel), and finally with aperture size (right panel). For the aperture size tests, we generated all galaxies with \textit{r}=15 mag and $R_{\textrm{eff}}$=5$\arcsec$, to keep the total flux, and hence the total SNR, roughly constant and ensure only the aperture changes affect the result. In all panels, the solid diagonal line shows a 1-to-1 relationship, or ideal performance compared to the baseline image. 

For each series of tests, we fit the relationship between the observed and the baseline $\acas{}$ using two models: an exponential and a linear offset, shown below.

\begin{align*}
    \rm{[Model~1]}\quad \acas{}_{\rm{obs}} &= \acas{}_{\rm{base}} \exp \left[ - \left( \frac{p}{\tau_p}\right)^{\alpha_p} \right] \\
    \rm{[Model~2]}\quad \acas{}_{\rm{obs}} &= \acas{}_{\rm{base}} - m_p \times p
\end{align*}

\noindent where $p$ is some variable (\avgsnr{}, PSF FWHM, or aperture size), and $\tau$, $\alpha$, and $m$ are parameters to be fit for each variable. We adopt the fit with the lowest root-mean-squared error divided by the degrees of freedom. 

\new{This is a similar approach to \cite{Thorp2021}, except our model is either a linear fit or an exponential rather than polynomial. An exponential model is a more natural fit than a polynomial, as it provides better-behaved constraints to the measured asymmetry, ensuring that it is bound between 0 and the baseline value, while the polynomial fit does not guarantee that the corrected asymmetry stays within the allowed range of $\acas{}$. Moreover, the free parameters of the exponential fit are intuitive: $r_p$ is a characteristic value of the parameter where the observed $\acas{}$ decreases significantly, and $\alpha_p$ is the strength of the dependence on $p$. In this exponential form, the free parameters are easy to interpret and readjust if needed. On the other hand, in high-degree polynomial fits, the meaning and the behaviour of the coefficients are less clear, and small adjustments can lead to dramatic changes in the fitted function. The simple straight-line fit in Model 2 does not have the same advantages of being bound between the baseline $\acas{}$ and 0; however, it is the most stable fit given it has only one free parameter. While the exponential model can provide good fits to all of our tests, we opted for the linear fits where adding an extra parameter did not give a significant improvement to the fit.} We discuss the trends seen in each panel below, and then show how the mathematical formula for $\acas{}$ gives rise to these trends in Sec. \ref{sec:cas_maths}.

\textbf{Signal-to-noise ratio.} The left panel shows how the observed $\acas{}$ changes with \avgsnr{}, while the resolution is fixed to 0.1$\arcsec$, same as the baseline image. The color gradient indicates decreasing \avgsnr{} from $>10$ (dark blue) to $1$ (light green). Points with \avgsnr{}$<1$ are colored in grey. 

The best-fitting model is the exponential decay:

\begin{equation}
\label{eq:a_cas_snr}
    \acas{}_{\rm{obs}} = \acas{}_{\rm{base}} \exp \left[ - \left( \frac{\langle \rm{SNR}\rangle }{0.74}\right)^{-0.89}\right]
\end{equation}

\noindent so that the observed $\acas$ approaches 0 as \avgsnr{} decreases, and approaches the baseline value as \avgsnr{} grows to infinity. We show the resulting lines of best fit for five different \avgsnr{} levels in Fig. \ref{fig:a_cas}. Our results are qualitatively consistent with previous results \citep[e.g.,][]{Bottrell2019,Thorp2021}: measured $\acas{}$ decreases with \avgsnr{}.

Most crucial is the fact that the bias between the measured and baseline $\acas{}$ depends not only on \avgsnr{} but also on the intrinsic asymmetry: galaxies with higher intrinsic $\acas{}$ suffer a larger measurement bias. Therefore, the dynamic range of $\acas{}$ in shallow imaging is low, making it more difficult to distinguish between disturbed and regular galaxies. For example, for SDSS-like imaging (\avgsnr{}$\sim$1) a measured $\acas{}=0.2$ maps to a wide range of intrinsic asymmetries between $0.2 < \acas{}_{\textrm{true}} < 0.8$, making it impossible to distinguish irregular and normal galaxies. Therefore, it is important to recognize that both the $\acas{}$ threshold to select disturbed objects must change for different datasets, and that we may only be able to detect extremely asymmetric galaxies in shallow imaging. While a correction can be applied, the scatter is also larger at lower \avgsnr{} values, making the correction less reliable in noisy images.

\textbf{Resolution.} The middle panel in Fig.~\ref{fig:a_cas} shows the dependence of $\acas{}$ on resolution, or PSF FWHM, in mock images with a fixed sky surface brightness (SB) of 30 mag/arcsec$^2$. Measured asymmetry decreases with resolution, as was previously found in other studies \citep{Lotz2004,Giese2016,Bignone2020,Thorp2021}, although the dependence is weaker than the dependence on \avgsnr{}. The best fit to our data is a small linear offset:

\begin{equation}
    \acas{}_{\rm{obs}} = \acas{}_{\rm{base}} - 0.038 \times \textrm{PSF FWHM}
\end{equation}

Interestingly, \cite{Thorp2021} report that the resolution-dependent error in $\acas{}$ is stronger and also depends on the \textit{intrinsic} asymmetry of the object, but we do not find this bias. This difference may be due to a difference in our simulated data. While \cite{Thorp2021} use realistic galaxies drawn from Illustris TNG, we use simplistic S\'ersic+Gaussian models described in Sec. \ref{sec:sim}. In our set of simulations, since the size and brightness of clumps are drawn from a random distribution, a galaxy is equally likely to have large-scale and small-scale disturbances. In other words, it is possible to have a galaxy with high intrinsic $\acas{}$ with only large clumps, or only small ones, or some fraction of both. Moreover, our clumps are circular, so we do not have asymmetric features that are large along one axis and small (or unresolved) along another, such as spiral arms. On the other hand, this is not true for realistic galaxies. First, real galaxies have spiral arms, that may contribute to asymmetry if perturbed. Galaxies with high intrinsic asymmetry may also have preferentially more small-scale features than symmetric galaxies, due to the physical difference between the two populations. For example, passively evolving galaxies will have a low $\acas{}$ and no small-scale star-forming clumps, while star-forming galaxies will typically have larger $\acas{}$ and more clumps, and large-scale features (spiral arms) that correlate with small-scale ones (star-forming regions). Mergers lead to both large $\acas{}$ and can induce star formation \citep[e.g.,][]{Ellison2008}. Thus if the relative number of small, unresolved features correlates with the overall $\acas{}$, then measured asymmetry will depend both on the resolution and intrinsic asymmetry. However, this is due to this secondary correlation rather than the numerical performance of the metric.

\textbf{Dependence on aperture size.} Finally, the right-most panel in Fig. \ref{fig:a_cas} shows the measured $\acas{}$ for a series of tests where galaxies were generated with a similar total SNR, a fixed sky background level of 23 mag/arcsec$^2$, and a resolution of 0.1$\arcsec$. We randomly drew an aperture size, ranging between 1$R_p$ and 2.5$R_p$, and calculated both the baseline and observed $\acas{}$ in that aperture to see how the size affects the measurement. Increasing the aperture leads to a larger underestimate of the baselie $\acas{}$. The best fit to the aperture dependence is a linear offset:

\begin{equation}
    \acas{}_{\rm{obs}} = \acas{}_{\rm{base}} - 0.12 \times fR_p
\end{equation}

\noindent Note that the aperture dependence incorporates two effects: the dependence of the measurement on the \textit{total} SNR (i.e., total flux of the galaxy), and on \avgsnr{}. Since all galaxies generated in this test have similar total fluxes and hence total SNR, the difference seen here is entirely due to changing the effective \avgsnr{} as the aperture size increases and includes more background flux. However, even for small apertures, there is still an offset from the diagonal due to the low total SNR, the effect seen in the left panel of Fig. \ref{fig:a_cas}. While we provide the fit to the linear offset here, the real correction must be applied by recalculating the \avgsnr{} for each object using Eq. \ref{eq:a_cas_snr}.

Ideally, if the background correction performs as expected, then adding more background pixels should not affect the final asymmetry, since their contribution is subtracted. However, as is clear in the panel, the bias depends linearly on the aperture size, with larger apertures being further away from the baseline measurement. This is important to keep in mind, since aperture size will depend both on the code implementation and the way $R_p$ is estimated.

\subsection{The problem is absolute}\label{sec:cas_maths}

The reasons for the noise and resolution dependence of $\acas{}$ stem from its mathematical definition, as previously mentioned in \cite{Thorp2021}. Here we expand on this discussion by considering the \textit{true} intrinsic CAS asymmetry of a galaxy, before an observation is taken -- the ideal noiseless measurement we would like to recover. A galaxy can be described by some surface brightness distribution, $\mathcal{I}(x,y)$, so its true asymmetry is:

\begin{equation}
        \acas{}_{\textrm{true}} = \min_{\mathbf{x_0}} \frac{\int_{\mathbb{R}^2} |\mathcal{I} - \mathcal{I}^{180}| d\mathbf{x} }{\int_{\mathbb{R}^2} |\mathcal{I}| d\mathbf{x}}.
\end{equation}

In reality, all astronomical observations are discretized to some pixel scale, so we also discretize the intrinsic light distribution. Note that at this stage, subpixel information is lost, so this is no longer strictly a ``true'' asymmetry as it will underestimate the effect of small-scale disturbances. The discretized noiseless asymmetry is:

\begin{equation}
    \acas{}_{\textrm{noiseless}}  = \min_{\mathbf{x_0}} \frac{\sum_{ij} |\mathcal{I}_{ij} - \mathcal{I}^{180}_{ij}| }{\sum_{ij} |\mathcal{I}_{ij}|} .
    \label{eq:a_cas_true}
\end{equation}

In an observation, the light from the galaxy is overlayed on some back- and foreground flux $I_{sky} (x,y)$, which may include the Earth's sky, stars, other galaxies, or other sources. It is then blurred, or convolved, with a PSF $\lambda(x,y)$, discretized on the chip, and is subject to a Poisson noise, $\epsilon$. Therefore, the \textit{observed} intensity is:

\begin{equation}
I_{ij} = \Big(\lambda * \big( \mathcal{I}(x,y) +  I_{sky}(x,y) \big)\Big)_{ij} + \epsilon_{ij}
\end{equation}

\noindent if there are sufficiently many electron counts and the Poisson noise can be approximated as Gaussian, $\epsilon_{ij}$ is normally distributed with $\epsilon_{ij}\sim\mathcal{N} \left(0, I_{ij}^2\right) \sim\mathcal{N} \left(0, \big( (\lambda * \mathcal{I})_{ij} + (\lambda *  I_{sky})_{ij}\big)^2 \right)$.

In general, this is a difficult function to deal with. However, we can show that the noise and resolution dependencies arise even with the simplest possible assumptions. 

\subsubsection{Noise dependence}\label{sec:cas_snr}

In the best case scenario, we can assume that the background is flat and subtracted, and ignore the contribution from the source Poisson noise\footnote{In real observations, noise originates both from the sky and the source itself. Although we do not model the Poisson noise from the source here, we note that in simulated galaxies used in our tests, both the baseline and the mock image have a Poisson noise contribution, but the Poisson noise does not significantly affect our results. This may be different for observations with low photon counts, such as ultraviolet or X-ray ones.}, so that the noise is effectively a Gaussian centered at 0 with a variance $\sigma^2_{sky}$. In this section, we also ignore the effects of the PSF by setting $\lambda=1$. We can then write the observed distribution simply as:

\begin{equation}
    I_{ij} = \mathcal{I}_{ij} + \epsilon_{ij}
\end{equation}

\noindent where $\epsilon_{ij} \sim \mathcal{N}(0, \sigma^2_{sky})$. 

We can now substitute this flux distribution in Eq. \ref{eq:a_cas} to calculate $\acas{}$. Ignoring the minimization step, we get:

\begin{align}
    \acas{} &= \frac{\sum_{ij} | (\mathcal{I}_{ij} - \mathcal{I}_{ij}^{180}) +
    (\epsilon_{ij} - \epsilon_{ij}^{180}) | }{ \sum_{ij} | \mathcal{I}_{ij} + \epsilon_{ij} | } - \langle \acas{}_{,bg}| \rangle.
\end{align}

\noindent Using the typical implementation of the $\langle \acas{}_{,bg} \rangle$ term (Eq. \ref{eq:a_cas_bg}), and omitting the aperture normalization term $N_{\rm{ap}}/N_{\rm{sky}}$ for clarity, the calculation is then:

\begin{align}
    \acas{} &= \frac{\sum_{ij} | (\mathcal{I}_{ij} - \mathcal{I}_{ij}^{180}) +
    (\epsilon_{ij} - \epsilon_{ij}^{180}) | - \sum_{ij}  | \epsilon_{ij} - \epsilon_{ij}^{180} | }{ \sum_{ij} | \mathcal{I}_{ij} + \epsilon_{ij} | }. 
\end{align}

An ideal measurement should recover the noiseless asymmetry in Eq. \ref{eq:a_cas_true}. However, the use of an absolute-valued function means that the equation above does not reduce to Eq. \ref{eq:a_cas_true}, even with our simplified assumptions. As pointed out in \cite{Thorp2021}, using the triangle inequality ($|a+b|-|b| \leq \acas{}$) it is apparent the numerator underestimates the true asymmetry. Moreover, the denominator (typically) is not corrected for the noise contribution at all.

Assuming that the galaxy is much brighter than the sky (i.e. in the high \avgsnr{} regime), the denominator can be approximated as $|\mathcal{I}_{ij} + \epsilon_{ij}| \approx |\mathcal{I}_{ij}| + |\epsilon_{ij}|$ and it may be possible to subtract the noise term. However, in the numerator, both the source residual $\mathcal{I} - \mathcal{I}^{180}$ and the noise residual $\epsilon_{ij} - \epsilon^{180}_{ij}$ may vary in signs while having similar magnitudes. If the underlying galaxy is bright but symmetric, it is possible to have $|\mathcal{I} - \mathcal{I}^{180}| \approx |\epsilon_{ij} - \epsilon^{180}_{ij}|$, making the sum inseparable. Therefore, even in the high-SNR regime, the calculation will not recover the true asymmetry in some cases.

In practice, this leads to a complex behaviour. In the low-\avgsnr{} regime, the noise contributes to both the numerator and the denominator, leading to over-subtraction of the noise with the $\acas{}_{,bg}$ term in the numerator, and the over-normalization in the denominator. This means that $\acas{}$ will produce a biased measurement, as seen in the left panel of Fig. \ref{fig:a_cas}, for low-\avgsnr{} galaxies. In the high-\avgsnr{} regime, the denominator approximates $\sum \mathcal{I}_{ij}$ and does not add a significant measurement error, while the measurement error in the numerator scales with the intrinsic asymmetry. In either case, the noise dependence of the $\acas{}$ measurement cannot be easily subtracted and correlates with the intrinsic asymmetry of the source.

\begin{figure}
    \centering
    \includegraphics[width=\linewidth]{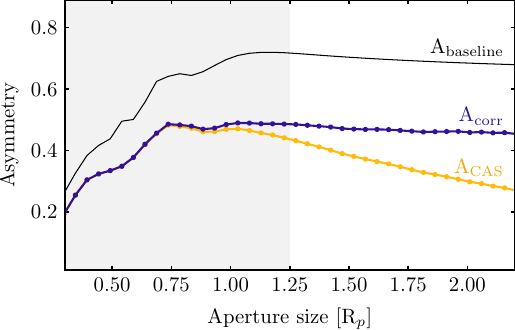}
    \caption{The dependence of $\acas{}$ measurements on aperture size for one simulated galaxy ($n$=3.3, \textit{r}-mag=14) with sky surface brightness of 22 mag/arcsec$^2$, showing the baseline $\acas{}$ (black), observed $\acas{}$ (yellow), and a corrected measurement $\acas{}_{\rm{corr}}$ (as defined in Sec. \ref{sec:a_corr}, purple). All asymmetric clumps in the \textsc{GalSim} run are placed within 1$R_p$, so the baseline asymmetry flattens out at $\sim$1.2$R_p$ (unshaded region). Beyond this radius, only the background noise contributes to the baseline $\acas{}$ measurement. $\acas{}_{\rm{corr}}$ also flattens out, indicating a better noise correction, while the observed $\acas{}$ continues to decrease as more sky pixels are included in the aperture. However, both $\acas{}$ and $\acas{}_{\rm{corr}}$ underestimate the baseline $\acas{}$.}
    \label{fig:a_apsize}
\end{figure}

\subsubsection{Dependence on the aperture}

An important consequence of the \avgsnr{} dependence is the dependence of $\acas{}$ on the aperture in which it is calculated. Since the denominator in Eq. \ref{eq:a_cas} is not corrected for the background noise, adding blank sky pixels increases the denominator. Therefore, $\acas{}$ measurement decreases with aperture size. We show this for one of the simulated galaxies in Fig. \ref{fig:a_apsize}. $\acas{}$ measured on the baseline and the observed images are shown in black and yellow, respectively. The purple line corresponds to the corrected $\acas{}$, which we discuss below in Sec. \ref{sec:a_corr}.

For apertures smaller than the galaxy (grey-shaded region in Fig.~\ref{fig:a_apsize}), all $\acas{}$ measurements increase with the aperture size as we include more asymmetric features. However, once the aperture exceeds the extent of the galaxy and includes additional sky pixels, the measurements should flatten out and converge on a single value. We see this for $\acas{}_{\textrm{baseline}}$, but not for the observed $\acas{}$. As more sky is included in the aperture, its contribution to the absolute flux increases. The residual term in the numerator corrects for this (albeit not fully), while the normalization term in the denominator of $\acas{}$ does not. Therefore, the denominator term increases as the aperture grows, leading to lower $\acas{}$ values. 

The aperture dependence is important to consider. Different software packages use different aperture sizes in which asymmetry is calculated (e.g., 1.5$R_p$ in \cite{statmorph} and the maximum radius in the $A_{\rm{CAS}}$ calculation in \cite{Pawlik2016}). Moreover, finding $R_p$ in itself is non-trivial, and so different packages may produce different radii, even if the fraction of $R_p$ is kept the same. As a result, catalogs produced with different algorithms will give systematically offset $\acas{}$ measurements on the same images.

\begin{figure}
    \centering
    \includegraphics[width=\linewidth]{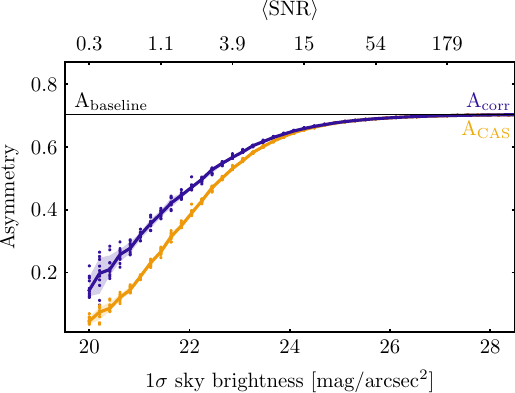}
    \caption{$\acas{}$ (yellow) and $\acas{}_{\textrm{corr}}$ (purple) as a function of sky brightness or \avgsnr{} for one simulated galaxy, compared to the baseline $\acas{}$ (black). Both measurements significantly underestimate $\acas{}$ up until \avgsnr{}$\sim$15, although this threshold will change depending on the intrinsic asymmetry of the object (see Fig. \ref{fig:a_cas}). The scatter in points represents 10 different noise-realizations of the mock image, the solid lines represent their median, and the shaded regions span the 16\textsuperscript{th} and the 84\textsuperscript{th} quantiles.}
    \label{fig:a_cas_corr_snr}
\end{figure}

\subsubsection{A noise correction}\label{sec:a_corr}

The noise dependence of $\acas{}$ has been noted since the parameter's introduction in \cite{Conselice2000}, and several solutions have been proposed. \cite{Giese2016} suggested using a library of simulated galaxies with known similar morphologies to correct for the bias in low-\avgsnr{} images. A similar approach would be to train a neural network on a set of deep or noise-free observations to mitigate observational biases \citep[e.g.,][]{Jia2024}. However, these solutions rely on assuming an intrinsic similarity between pairs or populations of galaxies, and therefore may disfavour finding rare objects with high intrinsic asymmetries. On the other hand, \cite{Thorp2021} had previously suggested correcting for the noise dependency numerically, by fitting a polynomial to measured $\acas{}$, intrinsic $\acas{}$, and \avgsnr{}. This approach does not rely on having any prior knowledge of the object, although the fit itself needs to be calibrated for different algorithms and possibly datasets. Similarly, we provide an exponential fit to the relationship between observed and baselined $\acas{}$ and \avgsnr{}; however, the scatter of this fit is larger for lower \avgsnr{} values. 

We suggest a third solution, which does not fully correct the measurement, but alleviates some of the noise dependence. We can re-define $\acas{}$ as

\begin{equation}
    \acas{}_{\textrm{corr}} = \min_{\mathbf{x_0}} \frac{\sum_{ij} |I_{ij} - I^{180}_{ij}|}{\sum_{ij} I_{ij}} - \langle \acas{}_{,bg} \rangle.
\end{equation}

\noindent where $\langle \acas{}_{,bg} \rangle$ is defined similarly to Eq. \ref{eq:a_cas_bg},

\begin{equation}
    \langle \acas{}_{,bg} \rangle = \frac{N_{\rm{ap}}}{N_{\rm{sky}}}\frac{\sum_{ij}^{sky} |\epsilon_{ij} - \epsilon^{180}_{ij}|}{\sum_{ij} I_{ij}}
\end{equation}

The only change from Eq. \ref{eq:a_cas} is the normalization term, where the residual flux is normalized by the total flux, rather than the total absolute flux. This definition had been previously suggested in \cite{Schade1995}, and used in practice \citep[e.g.,][]{Yagi2006,Shi2009,Wen2016,Bottrell2024}. With this definition, assuming the image is background-subtracted, the denominator is much more robust to changes in \avgsnr{} and does not require any noise corrections. This means that $\acas{}_{\textrm{corr}}$ is independent of the aperture size like the noiseless $\acas{}$, making this measurement more robust. The purple line in Fig. \ref{fig:a_apsize} shows that $\acas{}_{\textrm{corr}}$ flattens at the same radius as $\acas{}_{\rm{baseline}}$, as expected.

However, since the numerator is still affected by the noise flux, $\acas{}_{\textrm{corr}}$ will also produce biased asymmetry estimates. This is already seen in Fig. \ref{fig:a_apsize}, but we show it more clearly in Fig. \ref{fig:a_cas_corr_snr}. Here we plot the measured asymmetry for the observed $\acas{}$ and corrected $\acas{}$ (yellow and purple respectively) as a function of the sky brightness for one galaxy. While $\acas{}_{\rm{corr}}$ produces a less biased measurement than $\acas{}$, it is still significantly offset from the baseline value up to \avgsnr{}$>$15. Moreover, this \avgsnr{} threshold will change depending on intrinsic properties of the galaxy, as seein in Fig. \ref{fig:a_cas}. Thus, our noise-corrected $\acas{}_{\rm{corr}}$ is not a solution to the noise dependence, but at least a way to obtain consistent measurements regardless of the aperture size or $R_p$ definition on data with uniform depth.

\vfill\null
\subsubsection{Resolution dependence}\label{sec:a_cas_res}

The dependence on resolution stems from two factors. First, small-scale information is lost when it is discretized - any features smaller than the pixel scale cannot be recovered. Second, $\acas{}$ also depends on the PSF size.

Let us consider a noiseless case, ignoring for the time being noise, discretization, and minimization. We can write the observed flux convolved with a PSF as $I(x) = \lambda * \mathcal{I}$. Then the observed asymmetry integral is

\begin{equation}
        \acas{} = \frac{\int_{\mathbb{R}^2} |\lambda * \mathcal{I} - (\lambda * \mathcal{I})^{180})| d\mathbf{x} }{\int_{\mathbb{R}^2} |\lambda * \mathcal{I}| d\mathbf{x}}.
\end{equation}

\noindent \new{The rotation of the convolved image about some center \acent{} is given by the reflected PSF convolved with the rotated image (using the time reversal and translational properties of convolution): $(\lambda * \mathcal{I})^{180} = \lambda(-\mathbf{x}) * \mathcal{I}^{180}$. In many cases, such as in ground-based optical imaging, the PSF is roughly circularly symmetric described as a one-dimensional Moffat or similar profile. Even in interferometric observations, while the PSF is often an elongated beam, it is still symmetric under a 180$\degree$ rotation. However, in general the PSF may not be symmetric, and so it is better to consider the convolved and rotated image as a whole in practical applications. Nevertheless, to show the resolution dependence even in a best-case scenario, we assume a PSF that is invariant under reflection, so that $\lambda(-\mathbf{x})=\lambda(\mathbf{x})$. Then we the integral simplifies to}

\begin{equation}
        \acas{} = \frac{\int_{\mathbb{R}^2} |\lambda * (\mathcal{I} - \mathcal{I}^{180})| d\mathbf{x} }{\int_{\mathbb{R}^2} |\lambda * \mathcal{I}| d\mathbf{x}}.
\end{equation}

Since $\lambda$ and $\mathcal{I}$ are positive for all $\mathbf{x}$, we can simplify the denominator using the fact that for any Lebesgue-integrable, stricly positive functions, $|f*g|=|f|*|g|$ and $\int_{\mathbb{R}^n} (|f|*|g|) = (\int_{\mathbb{R}^n} |f|) (\int_{\mathbb{R}^n} |g|)$. Since the integral of the PSF is 1, the denominator is invariant under convolution:

\begin{equation}
\int_{\mathbb{R}^2} |\lambda * \mathcal{I}| d\mathbf{x} = \left(\int_{\mathbb{R}^2} |\lambda| d\mathbf{x}\right) \left(\int_{\mathbb{R}^2} |\mathcal{I}| d\mathbf{x}\right) =\int_{\mathbb{R}^2} |\mathcal{I}| d\mathbf{x}.
\end{equation}

On the other hand, the residual in the numerator is not invariant, since $\mathcal{I}-\mathcal{I}^{180}$ can be both positive and negative. In this case, by the triangle inequality, we have $|f*g| \leq |f|*|g|$, so

\begin{align}
\int_{\mathbb{R}^2} |\lambda * (\mathcal{I}-\mathcal{I}^{180})| d\mathbf{x} 
 &\leq  \int_{\mathbb{R}^2} |\lambda| * |\mathcal{I}-\mathcal{I}^{180}| d\mathbf{x}\\
 & = \int_{\mathbb{R}^2} |\lambda|d\mathbf{x} \int_{\mathbb{R}^2} |\mathcal{I}-\mathcal{I}^{180}| d\mathbf{x} \\
 & = \int_{\mathbb{R}^2} |\mathcal{I}-\mathcal{I}^{180}| d\mathbf{x}.
\end{align}

Combining the two, we can see that the asymmetry is underestimated:

\begin{align}
        \acas{} &= \frac{\int_{\mathbb{R}^2} |\lambda * (\mathcal{I} - \mathcal{I}^{180})| d\mathbf{x} }{\int_{\mathbb{R}^2} |\lambda * \mathcal{I}| d\mathbf{x}} \\
        &= \frac{\int_{\mathbb{R}^2} |\lambda * (\mathcal{I} - \mathcal{I}^{180})| d\mathbf{x} }{\int_{\mathbb{R}^2} |\mathcal{I}| d\mathbf{x}} \\
        &\leq \frac{\int_{\mathbb{R}^2} |\mathcal{I}-\mathcal{I}^{180}| d\mathbf{x}}{\int_{\mathbb{R}^2} |\mathcal{I}| d\mathbf{x}} = \acas{}_{\textrm{true}}
\end{align}

\begin{figure*}
    \centering
    \includegraphics[width=\linewidth]{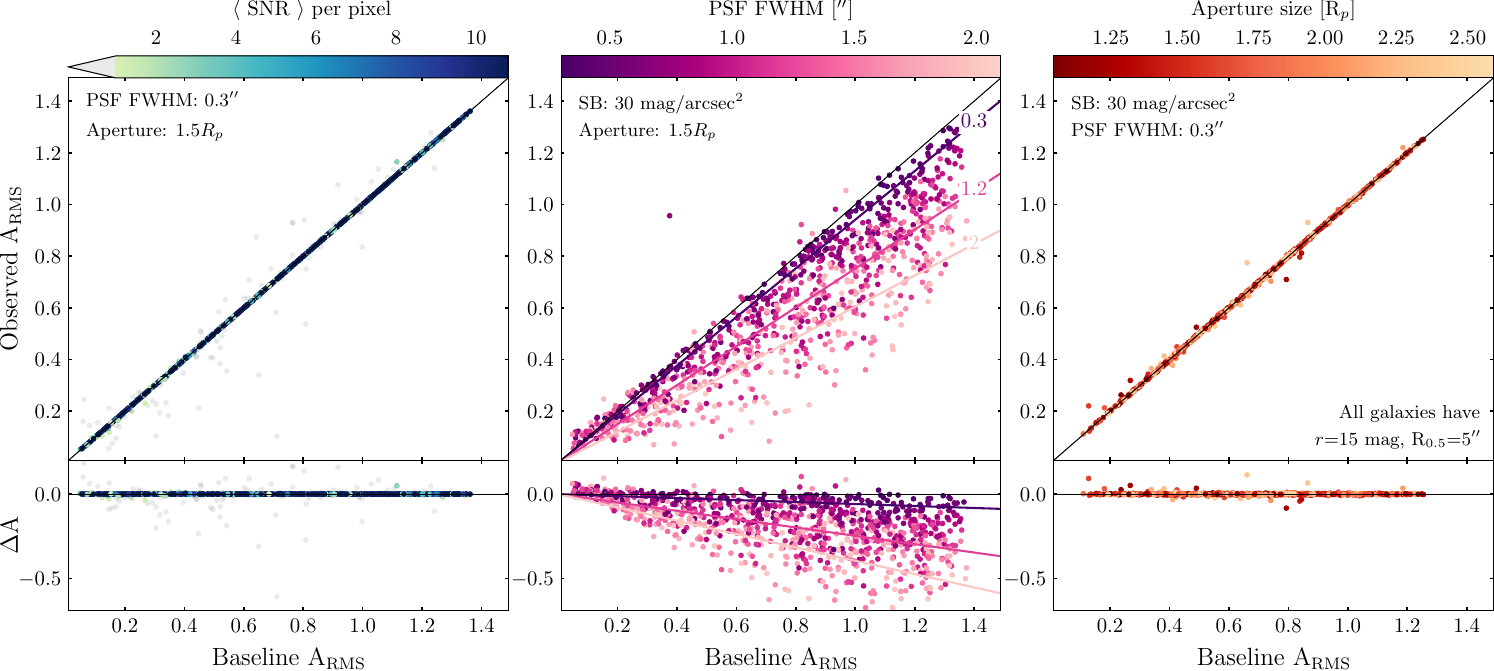}
    \caption{Same as Fig. \ref{fig:a_cas}, showing the test results for the $\arms$ measured on the mock observation \new{with Eq. \ref{eq:arms}} (y-axis) and the baseline image (x-axis). $\arms$ is independent of \avgsnr{} as long as \avgsnr{}$>1$. Similarly, since noise does not affect the measurement, $\arms$ is independent of the aperture size. However, $\arms$ has a stronger resolution dependence than $\acas{}$, significantly underestimating asymmetry for PSF FWHM $>$ 1$\arcsec$.}
    \label{fig:arms}
\end{figure*}

Thus, even ignoring the loss of sub-pixel information, the measured asymmetry systematically decreases due to the PSF. The magnitude of this effect would depend on the relative sizes of the PSF and the asymmetric features. It is possible to correct for the PSF dependence with deconvolution (see Sec. \ref{sec:fourier}), but the loss of the sub-pixel information is impossible to remedy analytically, although polynomial corrections may be tuned to particular observations \citep{Thorp2021}. One possible approach is to use machine learning and \textit{infer}, rather than measure, the small-scale asymmetry. Such ``super-resolution'' techniques have recently gained popularity \citep[e.g.,][]{Gan2021}, but it is difficult to train them to correctly predict properties of rare objects, which are often the target of galaxy evolution studies.

\section{RMS Asymmetry}\label{sec:arms}

RMS asymmetry, $\arms{}$, was first proposed in \cite{Conselice1997} and \cite{Conselice2000}, but did not become as widely used as $\acas{}$, since $\acas{}$ correlated better with galaxy color in the \cite{Conselice2000} sample.    However, as we showed in Sec. \ref{sec:cas}, the $\acas{}$ measurement is extremely sensitive to noise due to its absolute-valued formulation, and here we show that $\arms{}$ can alleviate this problem.

We define the true RMS asymmetry as:

\begin{align}
    \armssq{}_{\textrm{true}} &= \min_{\mathbf{x}_0} \frac{\int_{-\infty}^\infty (\mathcal{I} - \mathcal{I}^{180})^2 d\mathbf{x}  }{\int_{-\infty}^\infty \mathcal{I}^2 d\mathbf{x}}
\end{align}\label{eq:arms_true}

\noindent and once again discretizing the measurement, we get the noiseless (but resolution-dependent) RMS asymmetry is:

\begin{align}
     \armssq{}_{\textrm{noiseless}} &= \min_{\mathbf{x}_0} \frac{\sum_{ij} (\mathcal{I}_{ij} - \mathcal{I}^{180}_{ij})^2 }{\sum_{ij} \mathcal{I}^2_{ij} }.
\end{align}

\noindent Just like for $\acas{}$, we tested the response of $\arms{}$ to the changes in depth, pixel scale, and both, as shown left to right in Fig. \ref{fig:arms}. $\arms$ is invariant to changing \avgsnr{} down to \avgsnr{}$\sim 1$, and is independent of the aperture size. However, $\arms$ is more affected by changes in resolution - similarly to $\acas{}$, $\arms$ is underestimated in images with worse resolution in a way that depends both on the resolution, and the intrinsic asymmetry. 

\subsection{Robustness to noise}

For most \avgsnr{} regimes, $\arms$ perfectly recovers the true asymmetry. This is because the square of flux is mathematically better-behaved than the absolute value of flux, allowing us to separate out the source and noise contributions to the sum.

As before, let us assume that the noise is purely Gaussian and originates from the background, and ignore the effects of the PSF or minimization. Since $\mathcal{I} = I + \epsilon$, we can write the observed $\arms$ as

\begin{align}
    \armssq{}_{\textrm{obs}} &= \frac{\sum_{ij} \big(I_{ij} - I_{ij}^{180}\big)^2 }{\sum_{ij} I_{ij}^2 } \\
     &= \frac{\sum_{ij} \big((\mathcal{I}_{ij} - \mathcal{I}_{ij}^{180}) + (\epsilon_{ij} - \epsilon_{ij})\big)^2 }{\sum_{ij} (\mathcal{I}_{ij} + \epsilon_{ij})^2 }.
\end{align}

\noindent Since the source flux and the noise are uncorrelated in our approximation, and the noise is Gaussian with zero mean, $\langle \mathcal{I}_{ij} \epsilon_{ij} \rangle = 0$, so the cross-terms vanish and this equation simplifies to

\begin{equation}
    \armssq{}_{\textrm{obs}} = \frac{\sum_{ij} (\mathcal{I}_{ij} - \mathcal{I}_{ij}^{180})^2 + 2\sigma_{sky}^2 }{\sum_{ij} \mathcal{I}^2_{ij} + \sigma^2_{sky} }.
\end{equation}

\noindent Therefore, we can recover the true $\arms{}$ from the observed image as

\begin{equation}\label{eq:arms_noiseless}
    \armssq{}_{\textrm{noiseless}} = \frac{\sum_{ij} (I_{ij} - I_{ij}^{180})^2 - 2\sigma_{sky}^2 }{\sum_{ij} I^2_{ij} - \sigma^2_{sky} }.
\end{equation}

Of course, we have made a number of assumptions, most importantly that the background noise is flat and well-described by $\mathcal{N}(0, \sigma^2_{sky})$. In reality, we can compute the background contribution to asymmetry more robustly, as described in Appendix \ref{app:bg}. \new{In practice, we estimate the background with a sky aperture, and calculate $\arms$ as:}

\begin{equation}\label{eq:arms}
\new{
    \armssq{} = \frac{\sum_{ij} (I_{ij} - I_{ij}^{180})^2 - N_{ap} \langle A^2_{\textrm{RMS,sky}} \rangle }{\sum_{ij} I^2_{ij} - N_{\rm{ap}} \langle I^2_{\rm{sky}} \rangle }.
}
\end{equation}

\new{While it is possible to calculate $\arms{}$ using Eq. \ref{eq:arms_noiseless} and an estimate of the sky $\sigma$, this would ignore possible complications to the background, such as gradients, fore- and background sources, and complex noise structures. Instead, we suggest to estimate the background contribution from a sky aperture (e.g., see Appendix \ref{app:bg}) and calculate $\arms$ with Eq. \ref{eq:arms}, as we did in this study.} 

In any case, the important advantage of $\arms{}$ is that the background contribution to asymmetry is completely separable, and thus we can recover the baseline asymmetry, as seen in the first panel of Fig. \ref{fig:arms}. We do not fit any relationship between the observed and the baseline $\arms{}$, since it is clear from the plot that the asymmetry is recovered perfectly up until \avgsnr{} reaches 1.

\subsection{Resolution matching}\label{sec:fourier}

While $\arms{}$ is a very robust measurement for varying noise levels, it depends on the resolution more than $\acas{}$, as seen in the middle panel in Fig. \ref{fig:arms}. The reasons for this dependence are the same as for $\acas{}$, which we described in Sec. \ref{sec:a_cas_res}. As we decrease resolution, we are able to resolve fewer small-scale features of the image, and thus the overall asymmetry decreases. The impact of the PSF is stronger in the case of $\arms$, since it redistributes the flux away from the brightest regions, reducing the total sum of squared fluxes. 

The best-fit relationship between $\arms{}$ and the PSF size is exponential:

\begin{equation}
\arms{}_{\rm{obs}} = \arms{}_{\rm{base}} \exp \left[ - \left( \frac{\rm{PSF~FWHM}}{3.7}\right)^{-1.1}\right]
\end{equation}

The resolution dependence then has two roots: due to PSF smearing and due to loss of sub-pixel information. It is possible to remedy the PSF effect, but not the discretization one -- we cannot recover the lost information on individual galaxies. Therefore, it is important to recognize that both $\acas{}$ and $\arms$ are \textit{not scale-invariant}, and it is only meaningful to compare asymmetries calculated for a given physical size. 

\new{Previously, \cite{Deg2023} also studied the resolution dependence of a squared asymmetry metric, and found a generally similar pattern to us, where the squared asymmetry is more dependent on resolution than the absolute asymmetry. While they did not fit a function to the resolution dependence, they also find that squared asymmetry follows roughly an exponential curve, decreasing slowly when the resolution is high and then plunging when the object becomes less resolved. However, they find that squared asymmetry is fairly constant at five or more resolution elements per object (the equivalent of 2$\arcsec$ PSF in our case), whereas we see significant errors at this resolution. The reason for this discrepancy is likely the fact that measurement depends on the physical size of the asymmetric features. \cite{Deg2023} modelled their galaxies as smooth distributions with large-scale features induced by adding a first-order Fourier mode to the symmetric profiles. This means that the disturbances are roughly a quarter of the size of the galaxy model, and hence are well-resolved with five resolution elements, making asymmetry more stable at this or higher resolutions. The rest of the dependence seen in \cite{Deg2023} likely comes from the PSF redistributing the flux. If smaller-scale disturbances were introduced in their study, we expect they would see a stronger resolution dependence.}

While $\arms$ is inherently not scale-invariant, we can improve the calculation by matching the resolution of the observed image to the baseline resolution. While it is not possible to recover sub-pixel features, we can remove the effect of the PSF. To do this, we \new{match the PSF of our mock images to the baseline PSF. Since the pixel scale is defined in terms of the PSF, we also need to} rescale our mock images to the baseline pixel scale (0.1$\arcsec$) using linear interpolation. In practice, this scale should be defined using a physical size of features of interest in pc. Since our galaxies are simulated to match z$\approx$0.1 observations, this pixel scale maps 200 pc features. Note that although the mock observations are now sampled on a fine pixel scale, any information that was originally unresolved is still lost since interpolation cannot recover it. After rescaling the images, we deconvolve them.

In general, deconvolution is a complicated task. Since several intrinsic light distributions can produce the same image under convolution,  the inverse is ill-defined when the image is noisy. There are many proposed deconvolution algorithms, some simpler and some more complex. Among the current state-of-the-art solutions are: 1) using machine learning-based techniques to achieve super-resolution \citep[e.g.,][]{Gan2021} and 2) optimizing the deconvolved image numerically \citep[e.g.,][]{Shibuya2022,starred,starred2}. However, both approaches have limitations. Neural network techniques typically have to be trained on a large dataset (although pre-training or self-supervised approaches can remedy this), and generally underperform on rare objects, such as disturbed galaxies, making this method less suitable for detecting asymmetries. On the other hand, numerical optimization techniques do not rely on the training data, but take a long time to converge for individual galaxies and so are difficult to apply to large survey datasets. 

In this work, we opted for a simpler and faster approach to deconvolve images directly in Fourier space. We use a transform similar to Wiener deconvolution \citep{Wiener1949}, which combines the PSF deconvolution with a noise-damping term. Our approach differs in that we aim to recover a \textit{noisy} deconvolved image to avoid denoising artifacts, and it is described in detail in Appendix \ref{app:fourier}. Similarly to \cite{firedec}, instead of deconvolving the image completely, we deconvolve it down to a narrow Gaussian PSF (3 pixels FWHM) to ensure the final image is Nyquist-sampled. Our resolution-matched images are then produced in the following way: we rescale the image to the desired pixel scale (0.1$\arcsec$) using linear interpolation, and then apply the transformation in Fourier space to deconvolve the observed image and recover the Nyquist-sampled noisy source image.

\begin{figure}
    \centering
    \includegraphics[width=\linewidth]{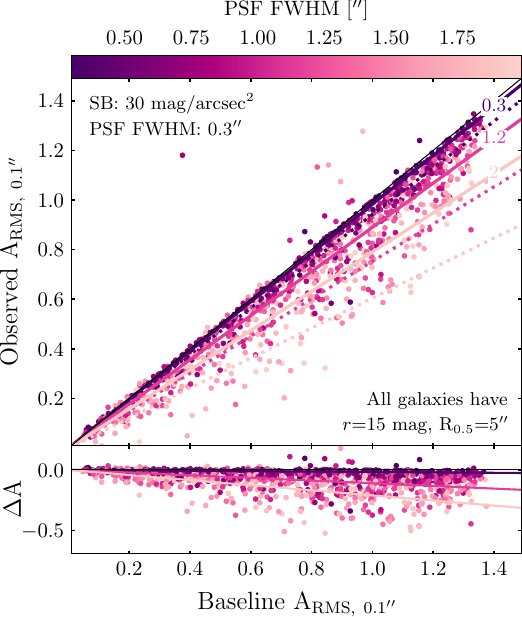}
    \caption{Same as the middle panel of Fig. \ref{fig:arms}, showing the resolution dependence of resolution-matched $\arms$. Here, the mock images were upscaled and deconvolved to match the baseline resolution of 0.1$\arcsec$. The solid lines show the best fit for the PSF size dependence for 0.3$\arcsec$, 1.2$\arcsec$, and 2$\arcsec$ PSFs. The dashed lines show the corresponding fits for the simple $\arms{}$ measurement without a resolution correction. While some resolution-dependent bias persists, it is reduced compared to the original $\arms{}$.}
    \label{fig:arms_corr}
\end{figure}

Fig. \ref{fig:arms_corr} shows the relationship between the baseline $\arms$ and the resolution-matched $\armsres$. We omitted the \avgsnr{} and the aperture size series from this plot since resolution matching does not affect the results. Again, the best fit for the dependence on the PSF size is exponential, but with a smaller offset for a given resolution:

\begin{equation}
    \arms{}_{\rm{obs}} = \arms{}_{\rm{base}} \exp \left[ - \left( \frac{\rm{PSF~FWHM}}{5.7}\right)^{-1.4}\right]
\end{equation}

\noindent we show the best fit lines for three different PSF sizes in Fig. \ref{fig:arms_corr} as solid lines, and the best fit lines for the uncorrected $\arms{}$ measurement as dashed lines of the same color.  As seen in the figure, the resolution correction  reduces the measurement bias, although $\arms$ is still underestimated when the mock PSF is much larger than the baseline. The offset from the baseline measurement is large for PSF FWHM $>2\arcsec$, five times wider than the best-case PSF. As we discussed, although deconvolution can remedy the impact of the PSF on the flux distribution, it cannot recover the subpixel information, so providing a complete correction without making any inferences about the underlying lost structure is impossible.

\begin{figure*}
    \centering    \includegraphics[width=\linewidth]{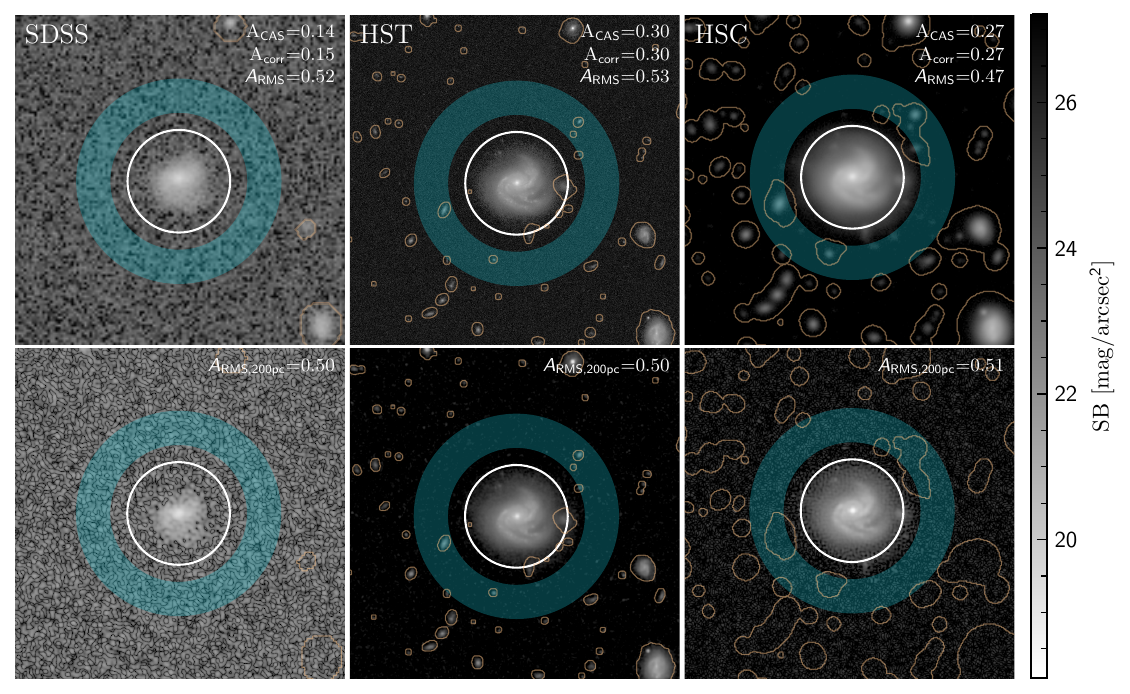}
    \caption{An example galaxy observed with SDSS (\textbf{left}), \textit{HST} (\textbf{center}), and HSC (\textbf{right}) in the \textit{I}-band filter or equivalent. The \textbf{top} row shows the pipeline-reduced images and $\acas{}$, $\acas{}_{\textrm{corr}}$, and $\arms$ measurements. Foreground objects were masked (thin orange contours), and all asymmetry metrics were calculated within 1.5$R_p$ (white solid circle). Sky asymmetry was estimated in a 2--3$R_p$ annulus (blue region). The \textbf{bottom} row shows resolution-matched images: SDSS and HSC observations were both upscaled and deconvolved down to 200 pc/px resolution, and the \textit{HST} image was downscaled and convolved with a 3-pixel-wide Gaussian PSF. The spiral structure is visibly sharper after deconvolution, and the deconvolved HSC image is visually similar to the \textit{HST} one. While both $\acas{}$ and $\arms$ measurements show an inconsistency between at least one pair of observations, $\armspc$ measurements are consistent between all three.}
    \label{fig:obs_example}
\end{figure*}

\section{A pilot test on observed data}\label{sec:obs}

In this section, we present the tests of all the asymmetry measurements against a pilot dataset of real observed. While simulations can show general trends in performance, they cannot test the robustness of our metric to different artifacts, non-Gaussian noise, or real asymmetric features, and so we turned to observational data.

For this test, we used the sample from \cite{Sazonova2021}, who studied disturbances in post-starburst galaxies at z$\approx$0.1. They constructed stellar mass and redshift-matched samples of quiescent, star-forming, and post-starburst galaxies that are observed with the HST and SDSS in the same filters. We selected 27 galaxies from the sample that also have \textit{i}-band HSC observations. While this is a small dataset, it spans a selection of mass- and redshift-matched star-forming, quiescent, and post-starburst / post-merger galaxies, and therefore a large range of intrinsic asymmetries. This dataset is an ideal test sample to evaluate the dependence of asymmetry measurements on both resolution and signal-to-noise ratio: it provides deep, high-resolution observations from \textit{HST}, deep but lower-resolution observations from HSC, and shallow low-resolution observations from SDSS. The observations were all taken in similar filters (\textit{I}-band or equivalent), so there are no intrinsic differences between observed galaxy structures. Therefore, a perfect $A$ measurement should produce the same result for a given galaxy for all three instruments. A table providing some basic properties of this sample (mass, redshift, and depths of each image) as well as all asymmetry measurement is provided in Appendix \ref{app:data}.

\subsection{Data processing}

Before calculating asymmetry, we needed to process the images. The SDSS observations were obtained from the SDSS Science Archive Server\footnote{\href{https://dr16.sdss.org}{SDSS DR16 Science Archive Server -- dr16.sdss.org}} \citep{sdss4}, and the HSC ones from the Image Cutout service of the HSC Subaru Strategic Program\footnote{\href{https://hsc-release.mtk.nao.ac.jp/doc/index.php/data-access__pdr3/}{HSC SSP Data Access}} \citep{hsc}. These observations were already reduced by their respective standard pipelines. The \textit{HST} data was obtained as part of a snapshot proposal (Proposal 14649; PI Alatalo), and reduced as described in \cite{Sazonova2021}. The images were used to create 40$\arcsec$ cutouts centered on each galaxy in the sample. An example galaxy observed with all three instruments is shown in Fig. \ref{fig:obs_example} (top row) with the SDSS image on the left, \textit{HST} in the center, and HSC on the right.

Next, we needed to mask all extraneous sources on the image and calculate the Petrosian radius. We opted to do this independently for each image. This choice can, in theory, add some differences in measured parameters: faint sources that are not masked in shallow observations can add to background asymmetry and thus affect our results. However, this is the most realistic set-up, as in most cases a deeper image is not available to construct the mask.

We followed the 3-step segmentation routine from \cite{Sazonova2021}. This algorithm uses \textsc{photutils} to construct three segmentation maps -- ``hot'', ``cool'', and ``cold'' -- and combine these maps into one map that aims to both capture faint envelope of the main source, and mask sources outside of the galaxy or bright foreground clumps within the galaxy. We grew the resulting mask by 5\% of the image size to capture any remaining faint contaminant light. Masked sources are shown with thin orange contours in Fig. \ref{fig:obs_example}. We then used \textsc{petrofit} \citep{petrofit} to calculate the Petrosian radius independently for each images. In Fig. \ref{fig:obs_example}, we show the 1.5$R_p$ region in which asymmetry is calculated as a solid white circle, and a sky region corresponding to 2--3$R_p$ as a blue annulus.

\begin{figure*}
    \centering
    \includegraphics[width=1.0\linewidth]{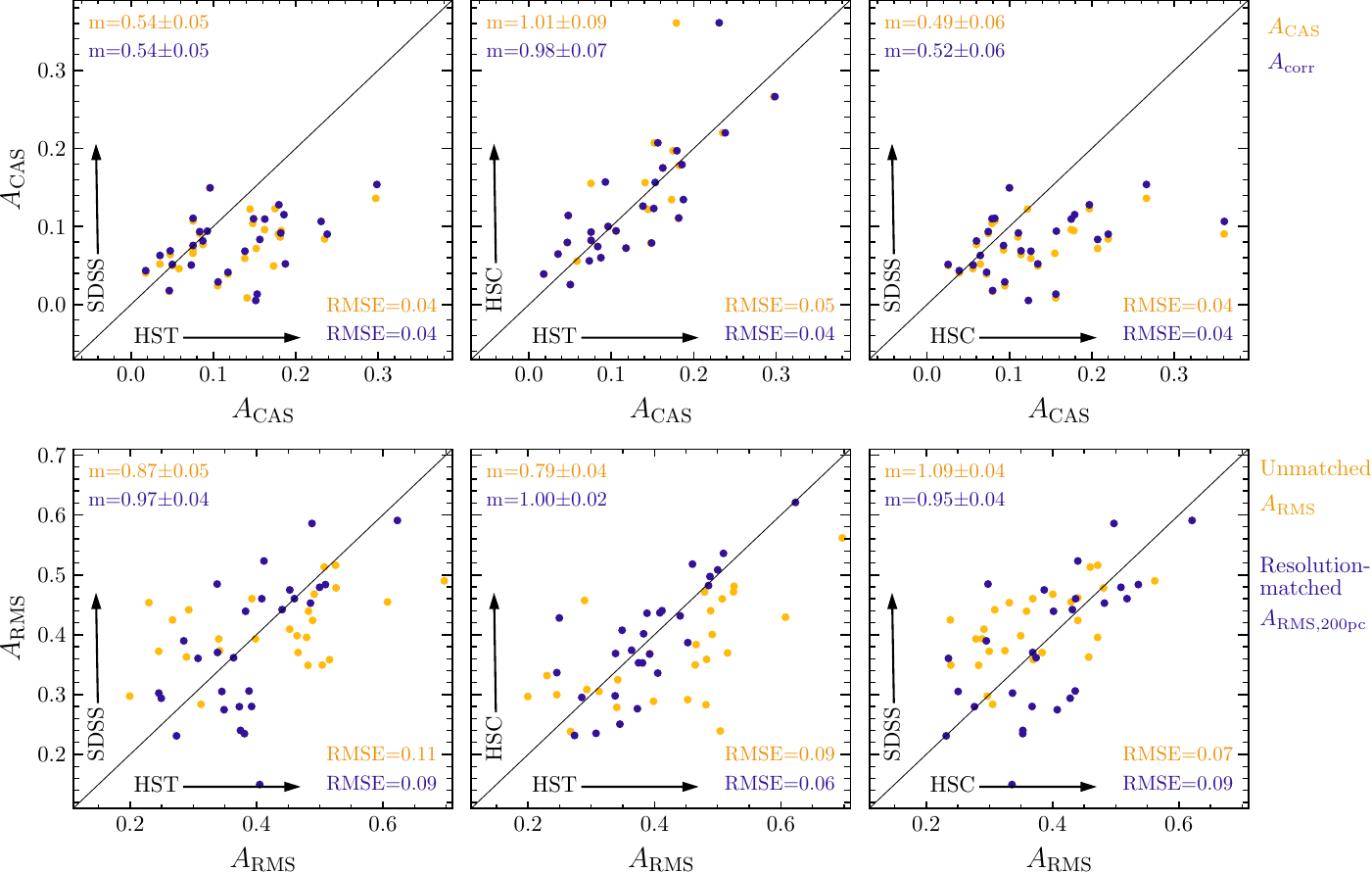}
    \caption{A comparison of asymmetry measured between pairs of observations of the same galaxies for $\acas{}$ metrics ($\acas{}$ and $\acas{}_{\textrm{corr}}$, \textbf{top}), and $\arms$ metrics ($\arms$ and $\armspc$, \textbf{bottom}). The comparisons are between SDSS and \textit{HST} (\textbf{left}), HSC and \textit{HST} (\textbf{center}), and SDSS and HSC (\textbf{right}). The diagonal line shows a one-to-one relationship, or an agreement between the metrics measured on two different images. The slope of the line of best fit for each sample is shown in the top-left, where a slope close to 1 indicates a good agreement and a shallower/steeper slope indicates a measurement bias. The root-mean-square error compared to the best fit is shown in the bottom right. $\acas{}$ is systematically underestimated in shallow images compared to deep ones (SDSS/\textit{HST} and SDSS/HSC), but is consistent in images of similar depth (HSC/\textit{HST}). Meanwhile, $\armsres$ provides consistent measurements for all three instruments, albeit with a large scatter around the diagonal, indicative of uncertainties in the asymmetry measurement.}
    \label{fig:a_observed}
\end{figure*}
Since we showed that both asymmetry measurements depend on resolution, we also calculate a resolution-matched $\armspc{}$ by rescaling all images to a pixel scale that corresponds to 200 pc per pixel. We do this by finding the angular resolution that corresponds to this physical resolution at the redshift of each object. The resulting pixel scale for each image is shown in Appendix \ref{app:data}. For all galaxies, this resolution is higher than either that of HSC and SDSS, so we deconvolved these images as described in Sec. \ref{sec:fourier}. On the other hand, \textit{HST} has a finer resolution, so we downscaled the \textit{HST} images to the required pixel scale and convolved them with a 3-pixel Gaussian PSF to ensure Nyquist sampling. While this means some of the information in the \textit{HST} images was lost, the $\armspc$ measurement requires a uniform spatial resolution, and we could not resample the shallow SDSS images any finer since resampling decreases \avgsnr{} below 1. In general, it is not feasible to upscale observations by more than a factor of a few, since both the \avgsnr{} decreases in the process, and smaller-scale features that are unresolved on the original pixel scale cannot be recovered anyway. Therefore, we opted for a matched resolution of 200~pc, which can capture asymmetries such as large star-forming regions, spiral arms, dust lanes, and tidal tails. The bottom panel of Fig. \ref{fig:obs_example} shows the resolution-matched \textit{HST}, HSC, and SDSS images. As seen in the image, our deconvolution routine (Appendix \ref{app:fourier}) sharpens the spiral structure in ground-based observations; most strikingly, the HSC image after deconvolution is visually very similar to the \textit{HST} image sampled on the same pixel scale. 

\subsection{Performance of asymmetry metrics}

For each galaxy and each instrument, we measured four asymmetries: $\acas{}$, corrected CAS asymmetry $\acas{}_{\textrm{corr}}$, RMS asymmetry $\arms$, and resolution-matched $\armspc$. The results for our sample are shown in Fig. \ref{fig:a_observed}. The figure is split as follows: we plotted $\acas{}$ and $\acas{}_{\textrm{corr}}$ in the top row (yellow and purple, respectively), while  $\arms$ and $\armspc$ are in the bottom row (yellow and purple). On each panel, we plotted asymmetry of each galaxy measured with one instrument on the x-axis and another on the y-axis. The solid black line indicates a 1-to-1 agreement between the instruments. For each set of measurements, we also calculated the best-fit gradient, computing the 1$\sigma$ uncertainty on the gradient via bootstrapping. A gradient close to one indicates that the measurements between two datasets are consistent, whereas a shallower or a steeper gradient indicates a bias.

Looking at the top row, we see again evidence for the dependence of $\acas{}$ on image depth. The measurements are consistent between HST and HSC observations (middle panel) -- which are both very deep. On the other hand, neither HST nor HSC agree with shallower SDSS observations. We also see a strong dependence on the intrinsic asymmetry: more disturbed galaxies have a higher measurement bias in SDSS, as seen by the shallow best-fit gradients. As a result, asymmetry measurements made using SDSS will have a much lower dynamic range, and thus are unable to differentiate truly undisturbed galaxies from weakly disturbed ones. $\acas{}_{\textrm{corr}}$ shows the same error as $\acas{}$, which is expected from our tests, since $\acas{}_{\textrm{corr}}$ corrects the dependence of $\acas{}$ on the aperture size but not overall \avgsnr{}-dependent bias (see Fig. \ref{fig:a_apsize}).

On the bottom row, we see a much better agreement between SDSS, \textit{HST}, and HSC. For the simple $\arms$ measurement, the results are consistent between SDSS and HSC (which have more comparable resolution), but there is a shallow slope indicating a bias comparing to \textit{HST} observations. On the other hand, for the resolution-matched $\armspc$ measurement, in all three cases the slope is consistent with one, indicating no bias when comparing the results between three different instruments. However, there is a significant scatter around the one-to-one line, with a root-mean-squared error of 0.06--0.09. The scatter represents various observational uncertainties that are not present in the simulated dataset: the difference in masking of foreground and background sources between images, artifacts that can affect the measurement, the slightly variable $R_p$, and the challenging nature of deconvolution. However, the scatter is not biased in either direction, making resolution-matched $\arms$ an overall more robust metric. Instead, we can interpret this scatter as the expected uncertainty of each asymmetry measurement when comparing catalogs made with different observational datasets, particularly where the native pixel scales are very different. 

\section{Discussion}\label{sec:discussion}

\subsection{Asymmetry \& noise}

In this work, we aimed to understand why asymmetry measurements depend so strongly on imaging quality. We studied how four metrics depend on noise and resolution: $\acas{}$, a noise-corrected $\acas{}_{\textrm{corr}}$, a root-mean-square $\arms$, and a resolution-matched $\armsres$ (or $\armspc$, where the resolution limit is defined in terms of a physical scale). 

\begin{figure*}
    \centering
    \includegraphics[width=\linewidth]{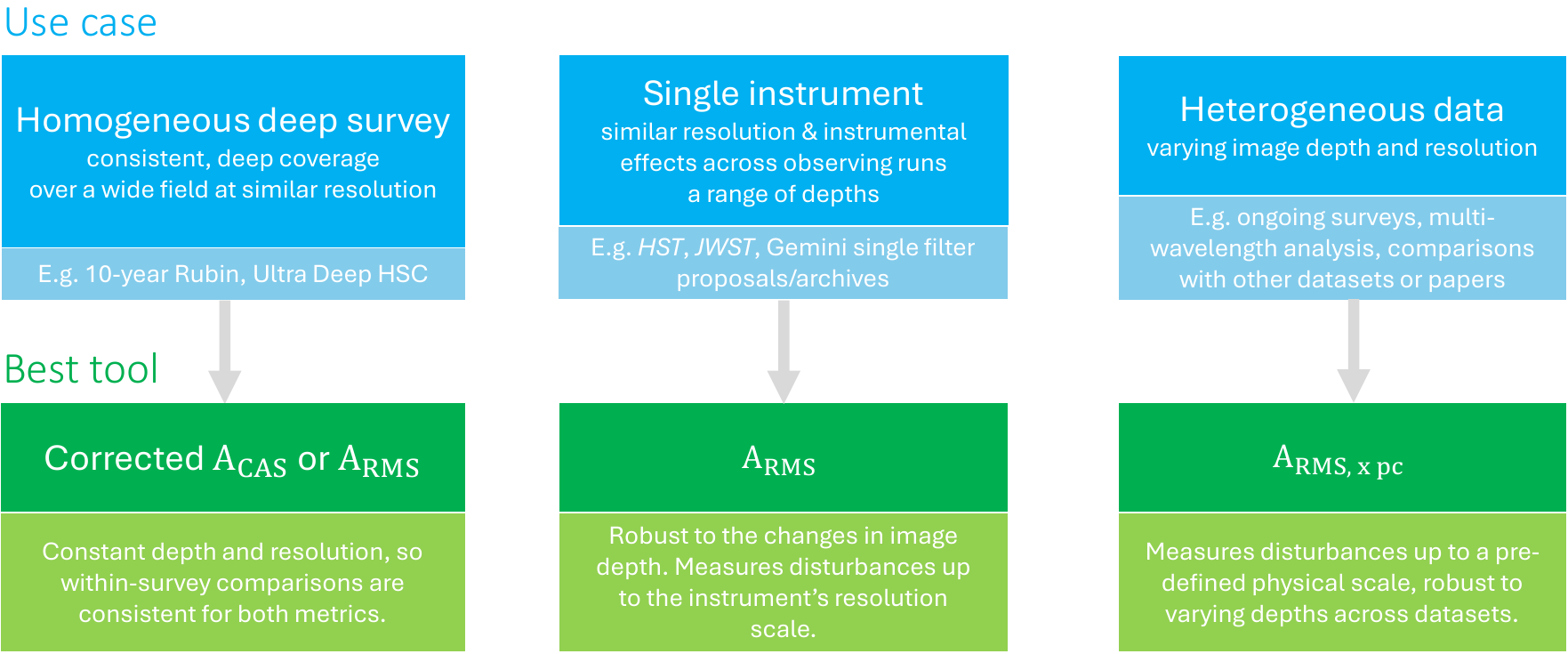}
    \caption{A schematic guidance on the the best asymmetry metrics for different studies, depending on the data type and origin. In general, $\armsxpc$ will provide the most robust measurements while enabling comparisons across different experimental or redshift regimes. }
    \label{fig:flowchart}
\end{figure*}

In Section \ref{sec:cas}, we showed that the dependence of $\acas{}$ on image depth arises due to a nonlinear dependence of the absolute-valued function on the noise term. In the extremely high signal-to-noise ratio regime, the noise contribution to the total flux is low, and therefore $\acas{}$ is a reliable metric, although even when \avgsnr{} is high, $\acas{}$ may be negative if the source is intrinsically very \textit{symmetric} (see Sec. \ref{sec:cas_snr}).

In shallower imaging, noise always contributes to the measurement despite the background asymmetry correction term. It adds bias both in the numerator of the equation (the asymmetry residual) and the denominator (the normalization flux), which leads to two issues: first, asymmetry is always systematically underestimated when noise is present; second, the measurement is highly dependent on aperture size as more sky pixels are added to the normalization flux. One easy correction that can be made is normalizing asymmetry with the total photometric flux, rather than the total absolute flux (using $\acas{}_{\rm{corr}}$, Sec. \ref{sec:a_corr}), thereby reducing the noise contribution to the normalizing flux. This correction remedies the aperture size and inclination dependence, but still provides biased estimates when \avgsnr{} is varied.

In Section \ref{sec:arms}, we studied a similar metric: a root-mean-square asymmetry, $\arms$. The idea of measuring asymmetry of the squared flux distribution was first introduced in \cite{Conselice2000} alongside the commonly used $\acas{}$, however it never gained popularity. At various times, it was re-proposed \citep[e.g.,][]{Deg2023}, but to this day an $\arms$ measurement has not gained traction. In this work, we looked at the behaviour of $\arms$ in more detail to compare it to $\acas{}$. We opted for a root-mean-square asymmetry, rather than a squared one, so that the strength of disturbances grows linearly rather than quadratically. 

We showed that $\arms$ is extremely robust to a changing signal-to-noise ratio, since the noise term easily decouples from the calculation in the assumption of uncorrelated Gaussian noise, above \avgsnr{}$>1$, and remains robust in realistic images with complex sky backgrounds (Section \ref{sec:obs}). The measurement scatter increases in extremely shallow imaging due to instabilities in the minimization step. We note that it may be possible to skip minimization (see Appendix \ref{app:min}), which we will explore in a future work. 

\subsection{Asymmetry \& resolution}

Both $\acas{}$ and $\arms$ are sensitive to image resolution, but this sensitivity is more extreme for $\arms$. This leads to a more philosophical discussion of the nature of the \textit{intrinsic} asymmetry of an object. Regardless of a computational strategy, asymmetry is defined as the difference between flux distributions under a rotation by 180$^\circ$. At infinite resolution, galaxies will possess many small-scale asymmetric features: star clusters, gas clouds, and even individual stars, leading to a high asymmetry value under this definition. This is a similar argument to that of measuring the length of a coastline \citep[e.g.,][]{Mandelbrot1967}: with increasing resolution levels, more irregular features appear, increasing the total length -- in that sense, galaxies, as a coastline, are fractal objects and thus a resolution-independent asymmetry value is impossible to define. Instead, just like with other fractal objects, it is useful to define a metric at a given resolution. Ideally, this resolution should be chosen as the physical scale of the disturbance we would like to detect.

In this work, we devised a new Fourier-based deconvolution approach to match the resolution of different observations (see Appendix \ref{app:fourier}). Our method is quick, but more robust methods exist -- for example, starlet deconvolution that uses optimization to find the best final image \citep{starred}. For the analysis of large survey data, we recommend using Fourier deconvolution as it is significantly faster, but optimization techniques should yield better results. 

Accounting for this resolution dependence, we defined a new measurement, $\armsx$ ($\armsxpc$), or RMS asymmetry measured at a spatial resolution of $x$ arcseconds ($x$ parsec). For the observed dataset, we calculated $\armspc$, RMS asymmetry resolved at 200~pc. At this resolution, we can detect large asymmetric features such as tidal tails, as well as disturbances to the spiral structure and relatively small satellite galaxies, but we do not resolve individual stellar clusters, so they do not contribute to the measured asymmetry. 

Our new metric is able to provide consistent asymmetry measurements independently of image depth, after the observations are resampled and (de-)convolved to match the desired resolution. However, there are two caveats. First, during upsampling, low-resolution observations lose effective \avgsnr{}. Therefore, with shallow observations such as that of SDSS, it is impossible to measure asymmetry on finer scales than those we tested in this work due to the loss of SNR. Our deconvolution approaches also dampens the noise-dominated high-frequency terms, so the effectiveness of deconvolution falls with decreasing \avgsnr{}. Second, it is impossible to recover features that are much smaller than the original pixel scale since upsampling simply interpolates between pixels -- it is thus important to find a balance between the desired resolution and one that the instrument is capable of reaching. We chose 200~pc for our tests as this required us to upsample SDSS images by a maximum factor of four, and allowed us to resolve structures where asymmetries can provide information about physical structures such as spiral arms and satellites.

The resolution dependence is especially important to keep in mind when comparing the structure of galaxies across different cosmic times. If galaxies observed in the same survey are located at different redshifts, the effective spatial resolution of these galaxies will change depending on their distance from us. Therefore, there will be a redshift-dependent bias in simple $\acas{}$ and $\arms{}$ measurements that may be misinterpreted as evolution of the ``disturbance'', when in fact, the disturbance of different features is probed at these different scales. Our new $\armsxpc$ measurement will probe the asymmetry of the same physical features, regardless of the object's redshift, and therefore allow a more consistent analysis.

However, care must still be taken when comparing galaxies of different physical sizes. \new{The overall size of the galaxy depends strongly on its mass \citep[e.g.,][]{Shen2003}, which can also affect the size of asymmetric features. Lower-mass galaxies will have smaller spiral arms and even smaller star clusters as the threshold for the Jeans instability depends on local density, which is generally lower in dwarfs.} Therefore, even a physically-motivated asymmetry may be probing different \textit{types} of features \new{in different galaxy populations}. However, not all features may have this size dependence: e.g., the size of the tidal tails depends also on the orbit and the stage of the merger \citep[e.g.,][]{Pawlik2018}. \new{An even more robust approach may be to first decide on the typical scale of asymmetries one wishes to detect given the aim of the study and the stellar mass range of the sample, and then match the images to that resolution. Thus, if one wanted to detect asymmetric tidal features, low-resolution imaging is sufficient (and increases effective signal to noise), while if one wanted to study the asymmetric spiral structure in galaxies, the required resolution limit will depend on the size of the galaxy. One solution is to look for a meaningful resolution limit in terms of the effective radius of the galaxy, as is often done in studies of molecular gas \citep[e.g.,][]{Deg2023}} This analysis is beyond the scope of this paper, but warrants further study.

\subsection{Detecting galaxy mergers}

One advantage of the $\armsxpc{}$ metric is that is independent of image properties, and provides consistent results at any redshift. While traditionally $\acas{} > 0.3$ has been used as a threshold above which a galaxy is considered ``disturbed'', this threshold will vary for datasets with different image properties. On the other hand, since $\armsxpc$ does not strongly depend on resolution or depth, such a threshold would be more universal, and we plan to determine the range of $\armsxpc$ typical of disturbed galaxies in a future study with a larger sample.

However, simply using a disturbance threshold to detect merging galaxies is not effective, since the observed morphology of a merging pair depends strongly on the merger stage, properties of the galaxies, and their orbits \citep[e.g.,][]{Pawlik2016}. Some mergers may have a detectable asymmetry while others may not: for example, late-stage mergers in early-type galaxies are seen as shells, an inherently symmetric feature. Selecting merger candidates using asymmetry alone will not provide either a pure or a complete sample \citep[e.g.,][]{Thompson2015,Bignone2017,Wilkinson2024}. 

If the goal is to obtain the most pure and/or complete merger sample, modern deep learning techniques ubiquitously outperform an ``asymmetry threshold'' approach \citep[e.g.,][]{Ackermann2018,Bottrell2019,DominguezSanchez2023}. The advantage of quantitative morphology metrics, however, is their interpretability and versatility. Different metrics probe different types of disturbances, and can be used in tandem to easily tailor the selection to a particular class of galaxies \citep[e.g., mergers of different stages;][]{Snyder2015a,Nevin2019}. To do this effectively, it is important to consider the types of features each metric is sensitive to.

Since both $\acas{}$ and $\arms{}$ are flux-weighted, they are more sensitive to bright disturbances rather than low-surface brightness features such as tidal tails. $\arms{}$ scales the residual flux quadratically, and so is even more biased towards bright pixels. Therefore, this metric is sensitive to internal asymmetries, such as star-forming clumps, double nuclei, or prominent dust lanes. This means that $\arms{}$ is best-suited for detecting mergers before the two cores coalesce. Moreover, the flux weighting means asymmetry may be better at detecting major mergers \citep{Nevin2019}, although \cite{Bottrell2024} show that a significant fraction of asymmetries are induced by minor and micro-mergers. On the other hand, shape asymmetry defined in \cite{Pawlik2016} is not flux-weighted, and so is sensitive to large-scale faint features, such as tidal tails. Combining $\arms{}$ and shape asymmetry will allow probing both the internal and tidally induced disturbances caused by the merger -- this will be tested on a large dataset in a future study.

Finally, asymmetry is a good metric to probe asymmetries that are unrelated to mergers, such as those caused by clumpy star formation and/or dust absorption \citep[e.g.,][]{Sazonova2021}; however, the relative magnitude of asymmetries induced by these features needs to be established on a large dataset.

\subsection{Considerations for surveys, pipelines, and catalogs}

While no measurements are able to perfectly reproduce the intrinsic asymmetry regardless of the imaging properties, they individually excel in different scenarios. We provide guidance on the performance of different metrics studied in this work in Fig. \ref{fig:flowchart}, to aid future analysis, survey, and pipeline design. 

In deep, homogeneous surveys, if the variations in image depth and resolution are minimal, both $\acas{}$ and $\arms$ can be used to obtain consistent asymmetry measurements within that survey area. However, in many surveys, depth can vary significantly \citep[e.g., in HSC COSMOS field, the depth varies by 2 magnitudes;][]{hsc}. If variations in seeing are less significant than variations in depth, $\arms$ will provide more consistent results. If using $\acas{}$, we recommend using the new corrected $\acas{}_{\textrm{corr}}$ with a non-absolute denominator term to reduce the noise-dependence. As shown in Sec. \ref{sec:a_corr}, the corrected $\acas{}$ measurement does not depend on aperture size, and therefore will be more robust to changing numerical implementations. Deep surveys such as the HSC Ultra Deep field or the upcoming Rubin LSST will both provide reliable $\acas{}$ and $\arms$ measurements. 

In shallow surveys $\arms$ will outperform $\acas{}$ since $\acas{}$ depends non-linearly on the intrinsic asymmetry. This means that the dynamic range of $\acas{}$ is small in surveys such as SDSS and Pan-STARRS, and only extremely disturbed  objects can be identified as asymmetric. For example, \cite{statmorph} show that most galaxies in the Pan-STARRS survey have $\acas{}<0.1$, with only a few galaxies identified as mergers. This low dynamic range can lead to us to underestimate the role of structural disturbances and mergers in galaxy evolution \citep[e.g.,][]{Sazonova2021,Wilkinson2024}.

For data obtained with a single instrument outside of a survey, e.g. as part of a proposal or an archival search, $\arms$ will be the more robust measurement. The resolution of a single space-based instrument is essentially constant while exposure times differ depending on the observational program, so $\arms$ is the better choice. For ground-based instruments, the seeing varies more, but the variations are still typically within 0.1$\sim$0.3$\arcsec$, while resulting depth can vary a lot more significantly. 

In general, astronomers would like to be able to compare our data to data obtained by other scientists, instruments, or surveys. Astronomical data are extremely heterogeneous both in terms of resolution and depth: it depends on the instrument, the filter, the weather conditions, the choice of the observing strategy, and more. Objects with different redshifts will necessarily have different physical resolution when observed with a single instrument. Therefore, for most applications, we recommend using $\armsxpc$ - RMS asymmetry at a given spatial resolution. Ideally, the spatial resolution should depend on the physical scale of the disturbances. In this work, we used $A_{\textrm{RMS,200pc}}$, which will be a good choice for most applications studying global structural evolution of galaxies, since it captures the largest star-forming regions, spiral structure, and tidal perturbations, while not accounting for the intrinsically asymmetric distribution of the individual star-forming clumps in the galaxies.

The implication of this new approach is that some high-resolution imaging, such as that obtained by \textit{HST} and \textit{JWST}, will need to be \textit{downgraded} to match the coarser resolution of $A_{\textrm{RMS,200pc}}$. While initially this sounds contentious, it is crucial to remember that the asymmetry metric is resolution-dependent by design, and so the physical scale of the features we wish to analyze needs to be set. With the resolution limit fixed, observers will still need to ensure that the observations they use for their measurements are rescaled to the corresponding pixel scale, which will depend on the redshift of the object. With these considerations, our advice for observers and pipeline developers who wish to provide the community with morphology catalogs that include asymmetry is the following:

\begin{enumerate}
    \item Compute both $\acas{}$ and $\arms$ on the reduced images, since these metrics are easiest to interpret and will be robust within the survey volume.
    \item Compute $\armsres$ on images rescaled to 0.1$\arcsec$, since this is a pixel scale that traces meaningful physical features for $z<2$ galaxies, and is attainable through deconvolution for most other surveys, including SDSS. This pixel scale is approximately the same as the native pixel scale of Euclid, Roman, and \textit{JWST}, and is easily achievable for LSST through upsampling by a factor of 2.
    \item Encourage users to use re-compute a physically-motivated $A_{\textrm{RMS,200pc}}$ measurement for their own datasets, especially when considering the evolution of asymmetry across cosmic time.
\end{enumerate}

Finally, some of the issues outlined here can be resolved with machine learning. For example, \cite{Desmons2023} and \cite{VegaFerrero2024} trained a model that can be used in the future to classify galaxies independently of the noise level by including noise augmentations during training, while \cite{Tohill2021} developed a neural network that predicts the CAS concentration, asymmetry, and smoothness simultaneously, giving a 10,000 speed-up compared to an analytical computation. Combining the two approaches could help mitigate the noise-dependence of $\acas{}$. However, neural networks perform worse on rare objects (such as mergers) as they are often under-represented in training sets, unless specifically tuned to be sensitive to those features, in which case they can over-predict the number of disturbed objects. Moreover, it is beneficial to use a metric that is initially defined to be better-behaved, such as $\armsxpc$, and then train a model to predict this metric in an automated way.

\section{Summary}

We have tested the performance of four asymmetry measurements: $\acas{}$ \citep{Conselice2000}, a corrected $\acas{}_{\rm{corr}}$ with a better normalization, root-mean-squared $\arms$ \citep{Conselice2000}, and $\armspc$ ($\armsres$) calculated on images with matched resolution of 200~pc (0.1$\arcsec$) per pixel. We evaluated these metrics on a set of simulated \textsc{GalSim} galaxies with varying image depth and resolution, as well as on a set of 27 galaxies observed with \textit{HST}, SDSS, and HSC in the \textit{I}-band. 

Our main results are:

\begin{enumerate}
    \item \textbf{$\acas{}$ severely underestimates asymmetry in shallow images} due to its absolute-valued definition (Sec. \ref{sec:cas_snr}). The bias grows exponentially with decreasing \avgsnr{}, and is large for images with SNR per pixel less than 5.\\
    Asymmetry of disturbed objects is affected more, leading to a lower overall dynamic range of $\acas{}$ in shallow surveys. Therefore, disturbed galaxies will have similar $\acas{}$ values to regular ones, making them difficult to identify.\\
    Another side effect of the noise impact is that $\acas{}$ depends on the aperture size. We provided a simple correction that alleviates the aperture size dependence (Sec. \ref{sec:a_corr}).

    \item \textbf{$\arms$ is insensitive to image depth} (Sec. \ref{sec:arms}). Due to defining $\arms$ as a square of the residuals, the background contribution can be easily subtracted, making this metric incredibly robust to changes in signal-to-noise ratio down to \avgsnr{}$\sim$1.

    \item \textbf{Asymmetry must be defined at a given physical scale.} Both $\acas{}$ and $\arms$ show resolution dependence (Sec. \ref{sec:a_cas_res} and \ref{sec:fourier}), although it is more significant for $\arms$. However, this is not something that can be easily corrected: at higher resolutions, as we are able to resolve smaller structures, all galaxies become increasingly asymmetric (Sec. \ref{sec:discussion}). Therefore, it is only meaningful to define asymmetry at a scale of features which constitute interesting ``disturbances''. We defined $\armsxpc$, i.e. $\arms$ measured at a given physical scale of $x$ pc, and implemented a method to rescale and deconvolve images down to the chosen scale (Sec. \ref{sec:fourier}). We tested its performance on observed data, where all images are rescaled to 200 parsec per pixel.

    \item \textbf{$A_{\textrm{RMS,200pc}}$ provides the best asymmetry measurements on observed data} (Sec. \ref{sec:obs}). When comparing  measurements obtained for the same galaxies using \textit{HST}, HSC, and SDSS imaging, only $\armspc$ results in unbiased measurements overall, while all other metrics underestimate asymmetry in SDSS (due to a lower depth), HSC (due to a lower resolution), or both. While the $\armspc$ measurement has a relatively large scatter when using SDSS data, there is no bias between the error of galaxies that are intrinsically regular or disturbed. The scatter is likely due to observational effects such as artifacts, different source masks, deconvolution effects, and different Petrosian radii.
\end{enumerate}

In Section $\ref{sec:discussion}$, we discussed our findings and provided guidelines for best use cases given our results. Overall, $\acas{}$ performs well in comparing data from deep surveys where \avgsnr{} is high. However, when comparing results from heterogeneous surveys where resolution and/or depth may vary, $\armsxpc$ is the most robust approach. 

This is important for designing morphological studies and pipelines for future surveys, as well as comparing morphologies obtained in different studies. The source SNR varies during survey duration, from one observing run to another, and with cosmic time. $\acas{}$ will result in inconsistent measurements in all of those cases. On the other hand, $\armsxpc$ will provide robust measurements across different surveys, epochs, and studies, as well as enable us to use archival data with potentially extremely variable resolution and depth. Therefore, our suggested guideline for survey and pipeline development is:

\begin{enumerate}
    \item Compute both $\acas{}$ and $A_{\rm{RMS}}$ on native survey images;
    \item Provide catalogs of $\armsres$ to allow easily comparison to other surveys;
    \item Encourage scientists to evaluate $\armsxpc$ for their samples of galaxies, especially if studying the redshift evolution of galaxy properties.
\end{enumerate}

The algorithms used to compute $\arms$ and $\armsxpc$, including the deconvolution step, will be provided as an open source package as part of the larger Rubin morphology pipeline (Sazonova et al., in prep.). \new{$\arms$ is also now included in the newest version of the \textsc{statmorph} package \citep{statmorph}. The \textsc{statmorph} $\arms{}$ algorithm differs slightly from the one used in this paper, namely in that the background and the center are estimated the same way for $\arms{}$ as for $\acas{}$ \citep[see][for details]{statmorph}. We also note that the \textsc{statmorph} implementation calculates $\arms$ \textit{squared}, and so we remind the users to take a square root of the output to get values consistent with those in this paper!}

\section*{Acknowledgements}
 
\new{We are very grateful to the anonymous referee for their fast and constructive feedback that helped us clarify important points in our paper, and to Peter Coles, the editor of the Open Journal for Astrophysics, with helping us during the submission process. We thank the members of the Rubin Galaxies Science Collaboration for their feedback on this work,} and the University of Waterloo for their support of the Canada Rubin Fellowship program. ES thanks Roan Haggar and Ana Ennis for many fruitful scientific discussions throughout this project. R.D. gratefully acknowledges support by the ANID BASAL project FB210003. D.D. acknowledges support from the National Science Center (NCN) grant SONATA (UMO-2020/39/D/ST9/00720). MF has been supported by the Polish National Agency for Academic Exchange (Bekker grant BPN/BEK/2023/1/00036/DEC/01), and acknowledges support from the Polish National Science Centre via the grant UMO-2022/47/D/ST9/00419. JR acknowledges funding from University of La Laguna through the Margarita Salas Program from the Spanish Ministry of Universities ref. UNI/551/2021-May 26, and under the EU Next Generation.

\vspace{5mm}
\textit{Facilities:} HST (WFC3, ACS, WFPC2); 2.5m Sloan Digital Sky Survey (SDSS) Telescope at Apache Point Observatory (APO), National Astronomical Observatory of Japan (NAOJ) 8.2m Subaru Telescope at Mauna Kea~Observatory.

\vspace{5mm}
\textit{Software:} \textsc{GalSim} \citep{galsim}, \texttt{statmorph} \citep{statmorph}, photutils \citep{photutils}, astropy \citep{astropy,astropy2,astropy3}, Scikit-Image \citep{skimage}, Matplotlib \citep{matplotlib}, pandas \citep{pandas}, NumPy \citep{numpy}, SciPy \citep{scipy}

\bibliography{references}{}
\bibliographystyle{mnras_custom}

\appendix

\section{Minimization}\label{app:min}

As defined in Eq. \ref{eq:a_cas} and Eq. \ref{eq:arms_true}, both asymmetry metrics require finding a center of rotation \acent{} such that asymmetry is minimized. 

Initially, asymmetry center was found through a grid search around the brightest pixel \citep[e.g.,][]{Conselice2000}; however, doing so for a sufficiently large search radius is computationally expensive. In modern applications, a numerical algorithm \citep[such as Nelder-Mead;][]{Nelder1965} is used to find a minimum of the asymmetry function. In this function, the image is first rotated using Scikit-Image \texttt{rotate} or equivalent, and then asymmetry is calculated. By default, \acent{} is allowed to vary continuously during minimization, meaning that rotation is performed about a fractional pixel. This requires interpolation, which adds correlations between background pixels in the rotated image that are not present in the original image. Because of these added correlations, the sky background of the original and the rotated images are no longer the same, and so calculating or even defining a background asymmetry term becomes challenging.

One solution is to not allow rotations about fractional pixel values \citep[or at most, allow half-pixel centers, as done in][]{Yagi2006}. However, minimization algorithms are not well-defined when the possible domain is discrete. Instead, we allowed rotation centers to vary continuously, but set the \texttt{rotate} function to perform 0\textsuperscript{th}-order interpolation, therefore avoiding any artifacts.

Another important step in the minimization is deciding on the initial guess for \acent{}. Since most algorithms perform a gradient-based search for a local minimum, starting with a poor guess for the global \acent{} can lead the minimizer towards a local minimum. A typical choice (e.g., in \texttt{statmorph}) is to use the first moment of the light distribution, or a flux-weighted center. This choice is actually a local \textit{maximum} of the asymmetry residual for most light distributions, and so a gradient-based minimizer may often converge on the wrong minimum.

Another concern is that minimization is the most computationally intensive step of the calculation, as it requires re-computing asymmetry for different \acent{} guesses. While traditionally optimization algorithms such as Nelder-Mead \citep{Nelder1965} are used, a significant boost in performance can be obtained by using gradient-based algorithms, or using tools that leverage automatic differentiation \citep[such as Jax;][]{jax}.

Finally, it may be possible to find the optimal center of the $\arms$ asymmetry without a computationally expensive minimization routine. While this is beyond the scope of this paper, we provide a quick proof-of-concept of this possibility.

Consider a noiseless, PSF-free, \textit{true} RMS asymmetry as defined in Eq. \ref{eq:arms_true}:

\begin{equation}
    \armssq{} = \min_{\mathbf{x}_0} \frac{\int_{-\infty}^\infty (\mathcal{I} - \mathcal{I}^{180})^2 d\mathbf{x}  }{\int_{-\infty}^\infty \mathcal{I}^2 d\mathbf{x}}.
\end{equation}

The integral in the denominator is simply a normalization term, and since it is independent of \acent{}, it can be omitted from optimization; so we only need to optimize the residual,

\begin{equation}
    R = \min_{\mathbf{x}_0} \int_{-\infty}^\infty (\mathcal{I} - \mathcal{I}^{180})^2 d\mathbf{x}.
\end{equation}

For this proof-of-concept exercise, we ignore that asymmetry is calculated within an aperture -- this does not affect the results in the noiseless case. Finally, for simplicity, we only consider this equation in one dimension; however, the calculations extend easily to multiple dimensions. 

We can define a rotation by 180$\degree$ as $\mathcal{I}^{180} = \mathcal{I}(x - 2x_0)$ for some rotation center $x_0$. Then we need to find \acent{} such that

\begin{equation}
    \frac{\partial R}{\partial x_0} = \frac{\partial}{\partial x_0} \int_{-\infty}^\infty (\mathcal{I}(x) - \mathcal{I}(x-2x_0))^2 dx = 0.
\end{equation}

Taking the partial derivative and expanding the square, we get:

\begin{align}
    0 &= \int_{-\infty}^\infty \mathcal{I}'(x-2x_0) \big[ \mathcal{I}(x) - \mathcal{I}(x-2x_0) \big] dx \\
    &= \int_{-\infty}^\infty \mathcal{I}'(x-2x_0) \mathcal{I}(x) dx - \int_{-\infty}^\infty \mathcal{I}'(u) \mathcal{I}(u) du \\
    &= \mathcal{I}\star\mathcal{I}'(-2x_0) - \frac{1}{2} \mathcal{I}^2(u) \Big|^{\infty}_{-\infty}.
\end{align}

Since the galaxies' flux vanishes at infinity, the second term is zero, therefore we need to find the zeros of the cross-correlation function between $\mathcal{I}$ and $\mathcal{I}'$. We can use the Fourier cross-correlation theorem:

\begin{figure*}
    \centering
    \includegraphics[width=\linewidth]{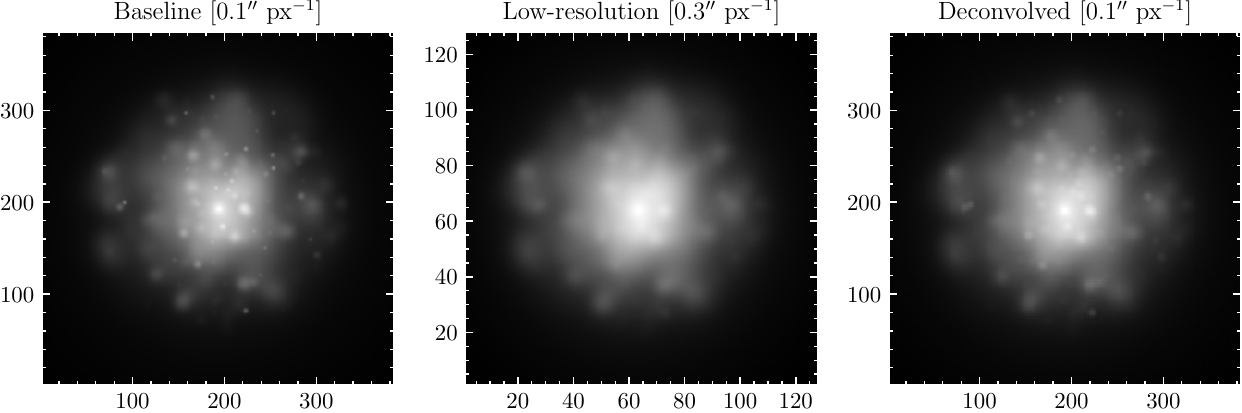}
    \caption{An example galaxy under deconvolution. \textbf{Left:} the baseline image generated on a 0.1$\arcsec$ pixel scale with a 0.3$\arcsec$ Gaussian PSF. \textbf{Middle:} a ``mock'' observation  generated on a 0.3$\arcsec$ pixel scale, with a 0.9$\arcsec$ Gaussian PSF, sky surface brightness of 30 mag/arcsec$^2$, and Poisson noise. \textbf{Right:} the mock image, after being resampled onto a 0.1$\arcsec$ px$^{-1}$ grid and deconvolved. Most small-scale features are recovered, although the smallest scales are lost as they cannot be recovered during interpolation. }
    \label{fig:deconv}
\end{figure*}

\begin{align}
    0 &= 
    \mathcal{I}\star\mathcal{I}'(-2x_0) = \mathcal{F} \Big[ \mathcal{I}\star\mathcal{I}'(2x_0) \Big] \\
    &= \frac{1}{2} \bar{\hat{\mathcal{I}}} \left(\frac{\xi}{2}\right) \hat{\mathcal{I}'} \left(\frac{\xi}{2}\right) =
    \frac{1}{2} \bar{\hat{\mathcal{I}}} \times \pi i \xi \hat{\mathcal{I}} \left( \frac{\xi}{2} \right) \\
    &= i\pi \xi \left| \hat{\mathcal{I}} \left( \frac{\xi}{2} \right) \right|^2.
\end{align}

Therefore, finding the theoretical asymmetry minimum of a noiseless distribution requires simply calculating the power spectrum of the image, and then finding a zero of this power spectrum -- although one would still need to find the zero that corresponds to the global minimum. While root-finding is still required, this is a much simpler problem to solve computationally than rotating the image and finding the asymmetry value at each rotation center. However, noise, discretization, and the PSF will affect this calculation. Determining whether an analytic \acent{} in a realistic observation exists is beyond the scope of this paper, but is an interesting avenue to explore.

\section{Background estimation}\label{app:bg}

Correctly estimating and subtracting the background contribution to asymmetry is crucial. This is a non-trivial task, since in theory, we need to estimate the contribution of the background that the source is overlayed on top of -- which is impossible without subtracting the source. In practice, many implementations will calculate the sky asymmetry in a blank patch of the sky, and extrapolate this to the expected asymmetry in the source aperture.

Different algorithms take slightly different approaches. For example, \texttt{statmorph} finds the largest possible \textit{blank} sky patch (i.e., a patch with no other sources found during the segmentation routine), and simply transpose the background to get the rotated sky. While this approach would give the correct background estimate for a purely Gaussian white noise sky, it will not be able to correct for possible large-scale fluctuations in the background.

Another approach is to calculate the background in \textit{several} sky patches randomly placed in the image away from the source \citep[e.g., in \textsc{Morfometryka};][]{Ferrari2015}. While similar to the \texttt{statmorph} algorithm, this approach is more robust against random fluctuations that may affect the background measurement in a single patch. Moreover, it is possible to then reconstruct the overall sky gradients and find large-scale asymmetries. 

In our tests, we opted to compute the sky background in annuli centered on the asymmetry center, similarly to the source asymmetry. This is the same approach as that used in typical photometry applications. There are several advantages to this method. First, we are able to capture any large-scale sky fluctuations. Second, the sky is rotated the same way as the image about the same center, therefore we are correctly estimating the sky contribution \textit{under the same rotation}. Finally, although we mask any extraneous sources both in the source and the sky aperture, any faint non-Gaussian sources (e.g., undetected stars or galaxies), will be included in our sky asymmetry estimate.

\section{Deconvolution}\label{app:fourier}

While in ideal noiseless observations deconvolution is a simple procedure, the addition of noise makes deconvolution an ill-posed problem, as noise terms are amplified by the inverse of the PSF. Iterative procedures, such as Richardson-Lucy deconvolution \citep{Richardson1972,Lucy1974} or newer methods \citep[e.g.,][]{starred} produce better results, but as they depend on numerical minimization, are computationally expensive. Instead, we turn to a simpler approach, motivated by the Wiener filter: deconvolving the image in Fourier space by penalizing noise-dominated high-frequency terms.

An observed distribution $I = \lambda * \mathcal{I} + \epsilon$ can be expressed in Fourier space as $\hat{I} = \hat{\lambda} \hat{\mathcal{I}} + \hat{\epsilon}$, where $\hat{\cdot}$ represents a Fourier transform. In theory, dividing by the PSF in Fourier space should recover the original distribution, $\mathcal{I}$. However, since $\hat{\lambda}^{-1} \rightarrow 0$ at high frequencies and Gaussian noise is constant at all frequencies, high-frequency noise terms are amplified.

The Wiener transform, $H_{W}$ \citep{Wiener1949}, aims to dampen the noise-dominated high-frequency terms while preserving the frequencies with high SNR. It is derived by minimizing the expected error between the deconvolved and the true images, $\langle |\hat{H}_W \hat{I} - \hat{\mathcal{I}} |^2 \rangle$. A Wiener transform acts essentially as a low-pass filter, setting high-frequency terms to 0. While this recovers the low-frequency features in the image, it can also add visible artifacts, since all high-frequency terms are removed. With these goals in mind, we want to find a trasnform $H$ that minimizes the error

Following the Wiener approach, we derive a new transform. The modifications of our method are two-fold: 1) first, we aim to recover the original noise level rather than denoise the image; and second, we do not fully deconvolve the image following the example in \cite{firedec} -- instead, we deconvolve it down to the Nyquist frequency. Both of these modifications help us remove denoising and Gibbs ringing artifacts.

Therefore, we want to recover a Nyquist-sampled and noisy image, accounting only for the smearing by the PSF: $\lambda_{3px} * \mathcal{I} + \epsilon$, where $\lambda_{3px}$ is a Gaussian PSF with a 3-pixel FWHM. We opted for retaining the noise term, since this reduces the artefacts introduced by denoising, and we have already shown that $\arms{}$ is independent of the SNR of the image. Therefore, we need to find a transform $H$ such that

\begin{equation}
    HI(x,y) = \lambda_{3px} * \mathcal{I} + \epsilon
\end{equation}

We can find $H$ by minimizing the expected difference between the right-hand and the left-hand sides of this equation, or

\begin{equation}
    \min_{H} \mathbb{E} \big| \hat{H} \hat{I} - (\hat{\lambda}_{3px} \hat{\mathcal{I}} + \hat{\epsilon})\big|^2
\end{equation}

Taking a partial derivative with respect to $\hat{H}$, we find that the optimal filter must take the form

\begin{equation}
    \hat{H}^* = \frac{
    \hat{\lambda} \hat{\lambda}_{3px}^*  + |\widehat{\textrm{SNR}}|^{-2}
    }{
    |\hat{\lambda}|^2 + |\widehat{\textrm{SNR}}|^{-2}
    }
\end{equation}

\noindent where $\widehat{\textrm{SNR}}$ is the signal-to-noise ratio term which damps the effect of the transform at noise-dominated frequencies. 

For a perfect deconvolution, $\widehat{\textrm{SNR}}$ must be equal to $\langle|\mathcal{I}|^2\rangle / \langle|\epsilon|^2\rangle$. However, this requires knowledge of the original distribution $\mathcal{I}$ and so is impossible to calculate directly. In photography, often one can assume that SNR decays with spatial frequency $\omega$ as $\widehat{\textrm{SNR}} \propto \omega^{-2}$. However, in general this is not true, and a careful estimate of $\widehat{\textrm{SNR}}$ is required for this deconvolution approach to work well. 

Instead of relying on any assumptions, we approximate $\widehat{\textrm{SNR}}$ from the observed images. We get the first order estimate by calculating it directly from the noisy image: $\widehat{\textrm{SNR}}' = \hat{I}/\hat{\epsilon}$. This calculation has two issues. First, as noise dominates the signal at high frequencies, $\hat{I}/\hat{\epsilon} \rightarrow 1$ while the true $\widehat{\textrm{SNR}} \rightarrow 0$. Therefore, we need some estimate of SNR in noise-dominated pixels. 

First, we mask all pixels with $\widehat{\textrm{SNR}}'$ below a threshold (80\textsuperscript{th} percentile or 3, whichever is higher), since we cannot use pixels where the noise contributes significantly to the ``signal'' measurement. Then, we find the 0\textsuperscript{th} frequency $\widehat{\textrm{SNR}}$, corresponding to the total SNR of the galaxy. We set the highest frequency $\widehat{\textrm{SNR}}$ to $10^4$ lower. We chose this value as it means that the light distribution spans roughly 10 magnitudes, consistent with the expected brightness difference between integrated light and the faint outer features. This produces a $\widehat{\textrm{SNR}}$ array where only the lowest- and the highest-frequency terms are populated: we then interpolate between these to fill the remaining values.

Second, while the interpolation mitigates the overestimation of $\widehat{\textrm{SNR}}$ at high frequencies, the resulting $\widehat{\textrm{SNR}}'$ effectively recovers $\hat{\lambda}\hat{I}/\hat{\epsilon}$ and thus underestimates the true $\widehat{\textrm{SNR}}$. Therefore, we further correct our $\widehat{\textrm{SNR}}'$ for the effect of the convolution by the PSF, estimating the intrinsic $\widehat{\textrm{SNR}}$ as:

\begin{equation}
    \widehat{\textrm{SNR}} = \frac{\widehat{\textrm{SNR}}'}{\hat{\lambda} + \widehat{\textrm{SNR}}'}
\end{equation} 

Fig. \ref{fig:deconv}c) shows the image deconvolved with our $H$. Unlike the Wiener deconvolution, our approach retains the original noise, thereby reducing the amount of artefacts that may affect the asymmetry measurement. We note that since the highest-frequency terms (e.g., point sources) are lost to the noise, it is impossible to recover them using our approach to deconvolution, and machine learning-based techniques are necessary to recover terms that are completely unresolved. Moreover, if computational time is not a constraint, more accurate optimization-based deconvolution methods may be used \citep[e.g.,][]{starred2}. However, our algorithms achieves reasonable performance in a fraction of the time required by optimization methods -- since the only bottleneck of this method is performing a fast Fourier transform, which takes less than a second for all but the largest image cutouts.
\onecolumngrid
\section{Observed sample} \label{app:data}

Here we provide the properties of the galaxies, instruments, and images used in the observational test in Sec. \ref{sec:obs}, as well as the measured asymmetry values for each galaxy. Refer to Table \ref{tab:sample} below.

{\renewcommand{\arraystretch}{1.5}
\begin{sidewaystable}[ht]
\centering
\label{tab:sample}
\caption{Properties of the observed dataset}
\begin{tabular}{cccc|ccccc|ccccc|ccccc}
\toprule

     \multicolumn{4}{c}{Galaxy properties} &
     \multicolumn{5}{c}{HST} &
     \multicolumn{5}{c}{HSC} &
     \multicolumn{5}{c}{SDSS}  \\
    Type &
    \colhead{$\log$ M$/$M$_\odot$} &
    \colhead{$z$} &
    \colhead{200~pc [$\arcsec$]} &
    \colhead{Depth} &
    \colhead{A$_{\rm{CAS}}$} &
    \colhead{A$_{\rm{corr}}$} &
    \colhead{A$_{\rm{RMS}}$} &
    \colhead{A$_{\rm{RMS}}^{200\rm{pc}}$} &
    \colhead{Depth} &
    \colhead{A$_{\rm{CAS}}$} &
    \colhead{A$_{\rm{corr}}$} &
    \colhead{A$_{\rm{RMS}}$} &
    \colhead{A$_{\rm{RMS}}^{200\rm{pc}}$} &
    \colhead{Depth} &
    \colhead{A$_{\rm{CAS}}$} &
    \colhead{A$_{\rm{corr}}$} &
    \colhead{A$_{\rm{RMS}}$} &
    \colhead{A$_{\rm{RMS}}^{200\rm{pc}}$} \\
\hline
  PSB & 10.8 & 0.139 & 0.079 & 23.7 & 0.18 & 0.23 & 0.70 & 0.62 & 26.2 & 0.36 & 0.36 & 0.56 & 0.62 & 23.7 & 0.09 & 0.11 & 0.49 & 0.59 \\
 PSB & 10.2 & 0.048 & 0.206 & 23.7 & 0.06 & 0.07 & 0.45 & 0.25 & 25.2 & 0.06 & 0.06 & 0.29 & 0.34 & 23.6 & 0.05 & 0.05 & 0.41 & 0.30 \\
 SFG & 10.2 & 0.047 & 0.209 & 26.1 & 0.18 & 0.18 & 0.46 & 0.36 & 25.2 & 0.11 & 0.11 & 0.35 & 0.37 & 23.8 & 0.09 & 0.09 & 0.40 & 0.36 \\
 PSB & 11.0 & 0.129 & 0.084 & 23.7 & 0.17 & 0.19 & 0.52 & 0.44 & 26.4 & 0.13 & 0.13 & 0.37 & 0.43 & 23.8 & 0.05 & 0.05 & 0.36 & 0.44 \\
 PSB & 10.2 & 0.033 & 0.294 & 23.7 & 0.10 & 0.10 & 0.39 & 0.34 & 26.4 & 0.10 & 0.10 & 0.37 & 0.30 & 23.8 & 0.15 & 0.15 & 0.46 & 0.48 \\
 QG & 10.4 & 0.092 & 0.113 & 25.8 & 0.04 & 0.04 & 0.25 & 0.31 & 27.9 & 0.06 & 0.06 & 0.30 & 0.24 & 23.6 & 0.05 & 0.06 & 0.37 & 0.36 \\
 SFG & 10.6 & 0.125 & 0.086 & 25.8 & 0.30 & 0.30 & 0.53 & 0.50 & 28.0 & 0.27 & 0.27 & 0.47 & 0.51 & 23.6 & 0.14 & 0.15 & 0.52 & 0.48 \\
 SFG & 10.6 & 0.119 & 0.090 & 25.8 & 0.08 & 0.08 & 0.34 & 0.34 & 28.0 & 0.09 & 0.09 & 0.32 & 0.37 & 23.6 & 0.07 & 0.08 & 0.37 & 0.37 \\
 QG & 10.9 & 0.125 & 0.086 & 25.7 & 0.14 & 0.14 & 0.50 & 0.41 & 27.9 & 0.13 & 0.13 & 0.24 & 0.34 & 23.8 & 0.06 & 0.07 & 0.35 & 0.15 \\
 SFG & 11.0 & 0.133 & 0.082 & 25.8 & 0.05 & 0.05 & 0.23 & 0.39 & 28.0 & 0.11 & 0.11 & 0.33 & 0.37 & 23.7 & 0.06 & 0.07 & 0.45 & 0.28 \\
 SFG & 10.9 & 0.165 & 0.068 & 25.8 & 0.11 & 0.11 & 0.34 & 0.28 & 28.1 & 0.09 & 0.09 & 0.28 & 0.30 & 23.7 & 0.02 & 0.03 & 0.39 & 0.39 \\
 QG & 10.3 & 0.080 & 0.128 & 25.8 & 0.09 & 0.09 & 0.29 & 0.38 & 28.1 & 0.06 & 0.06 & 0.46 & 0.35 & 23.8 & 0.08 & 0.08 & 0.36 & 0.23 \\
 QG & 10.7 & 0.104 & 0.102 & 25.8 & 0.15 & 0.15 & 0.48 & 0.39 & 28.1 & 0.08 & 0.08 & 0.36 & 0.44 & 23.8 & 0.10 & 0.11 & 0.44 & 0.31 \\
 QG & 10.8 & 0.092 & 0.113 & 25.8 & 0.12 & 0.12 & 0.48 & 0.37 & 28.1 & 0.07 & 0.07 & 0.28 & 0.35 & 23.6 & 0.04 & 0.04 & 0.35 & 0.24 \\
 SFG & 10.2 & 0.107 & 0.099 & 25.8 & 0.14 & 0.15 & 0.47 & 0.41 & 28.1 & 0.16 & 0.16 & 0.38 & 0.44 & 23.6 & 0.01 & 0.01 & 0.37 & 0.46 \\
 SFG & 10.5 & 0.076 & 0.135 & 25.8 & 0.16 & 0.16 & 0.61 & 0.46 & 28.0 & 0.17 & 0.17 & 0.43 & 0.52 & 23.7 & 0.10 & 0.11 & 0.45 & 0.46 \\
 SFG & 10.8 & 0.092 & 0.113 & 25.8 & 0.23 & 0.24 & 0.49 & 0.49 & 27.8 & 0.22 & 0.22 & 0.44 & 0.48 & 24.1 & 0.08 & 0.09 & 0.42 & 0.45 \\
 QG & 11.0 & 0.123 & 0.087 & 25.7 & 0.02 & 0.02 & 0.20 & 0.27 & 26.9 & 0.04 & 0.04 & 0.30 & 0.23 & 23.7 & 0.04 & 0.04 & 0.30 & 0.23 \\
 QG & 11.2 & 0.175 & 0.065 & 25.9 & 0.05 & 0.05 & 0.31 & 0.37 & 27.8 & 0.08 & 0.08 & 0.31 & 0.28 & 23.8 & 0.02 & 0.02 & 0.28 & 0.28 \\
 SFG & 10.2 & 0.063 & 0.161 & 25.3 & 0.15 & 0.16 & 0.49 & 0.41 & 27.7 & 0.21 & 0.21 & 0.40 & 0.44 & 23.7 & 0.07 & 0.08 & 0.47 & 0.52 \\
 QG & 10.9 & 0.087 & 0.119 & 25.1 & 0.05 & 0.05 & 0.40 & 0.35 & 25.6 & 0.03 & 0.03 & 0.29 & 0.25 & 23.6 & 0.05 & 0.05 & 0.39 & 0.31 \\
 SFG & 10.8 & 0.082 & 0.126 & 25.9 & 0.08 & 0.08 & 0.29 & 0.35 & 26.3 & 0.07 & 0.07 & 0.31 & 0.41 & 23.8 & 0.09 & 0.09 & 0.44 & 0.27 \\
 QG & 10.5 & 0.082 & 0.126 & 25.9 & 0.08 & 0.08 & 0.27 & 0.25 & 26.2 & 0.08 & 0.08 & 0.24 & 0.43 & 23.9 & 0.11 & 0.11 & 0.42 & 0.29 \\
 SFG & 10.2 & 0.028 & 0.340 & 25.1 & 0.17 & 0.18 & 0.84 & 0.45 & 25.7 & 0.20 & 0.20 & 0.44 & 0.39 & 23.9 & 0.12 & 0.13 & 0.46 & 0.47 \\
 QG & 10.4 & 0.066 & 0.153 & 25.2 & 0.08 & 0.09 & 0.53 & 0.51 & 26.8 & 0.16 & 0.16 & 0.48 & 0.54 & 23.9 & 0.07 & 0.09 & 0.48 & 0.48 \\
 SFG & 10.7 & 0.175 & 0.065 & 25.8 & 0.14 & 0.15 & 0.48 & 0.38 & 25.9 & 0.12 & 0.12 & 0.47 & 0.40 & 23.9 & 0.12 & 0.01 & 0.40 & 0.44 \\
 SFG & 10.9 & 0.188 & 0.062 & 25.9 & 0.18 & 0.19 & 0.51 & 0.49 & 25.9 & 0.18 & 0.18 & 0.46 & 0.50 & 23.8 & 0.09 & 0.12 & 0.51 & 0.59 \\
\bottomrule
\end{tabular}

\end{sidewaystable}
}

\label{lastpage}
\end{document}